\documentclass[12pt]{iopart}
\usepackage{amssymb}
\usepackage{amsfonts}
\usepackage{bm}

\newcommand{\nc}{\newcommand}
\nc{\JStP}{{\it J. Stat. Phys.}}  \nc{\IJMP}{{\it Intern. J. Mod. Phys.}}
\nc{\BBS}{{\rm BBS}\ }
\nc{\g}{\gamma}
\nc{\lm}{\lambda} \nc{\la}{\lambda}
\nc{\bh}{{\bf h}}
\nc{\cR}{{\cal R}} \nc{\kp}{{\varkappa}} \nc{\om}{\omega}   \nc{\CN}{{\cal N}}
\nc{\qt}{\tilde{q}} \nc{\tp}{\tilde{p}} \nc{\rt}{\tilde{r}}
\nc{\ty}{\tilde{y}} \nc{\tx}{\tilde{x}} \nc{\tQ}{{\widetilde Q}}
\nc{\trh}{\tilde{\rho}}
\nc{\ny}{\nonumber}
\nc{\lk}{\left(} \nc{\rk}{\right)} \nc{\Rb}{\right]} \nc{\Lb}{\left[}
\nc{\rb}{\right\}} \nc{\lb}{\left\{}
\nc{\hs}{\hspace*{1cm}} \nc{\hx}{\hspace*{3mm}} \nc{\hq}{\hspace*{6mm}}
\nc{\eR}{{\rm R}} \nc{\eL}{{\rm L}}
\nc{\CD}{{\cal D}}  \nc{\CDC}{\breve{\cal D}}  \nc{\WD}{{\widehat{\cal D}}}
\nc{\nul}{{(0)}} \nc{\one}{{(1)}}
\nc{\aab}{\boldsymbol{\alpha}}  \nc{\abb}{\boldsymbol{\beta}}
\nc{\pj}{{\textstyle \prod_{j\in\CD}}\;\qq{j}}
\nc{\half}{{\textstyle\frac{1}{2}}}

\nc{\al}{\alpha}  \nc{\sig}{\sigma}   \nc{\ZN}{\mathbb{Z}_N}
\nc{\bg}{\boldsymbol{\gamma}} \nc{\bdr}{{\boldsymbol{\rho}}}
\nc{\bu}{{\bf u}} \nc{\bv}{{\bf v}}
\nc{\bV}{{\bf V}}
\nc{\bnul}{{\mathbf 0}}
\nc{\bep}{\bu_{n+1}^{-1}(\ap-\bp\bv_{n+1})} \nc{\Xop}{\mathbf{X}}
\nc{\qq}[1]{e^{{\rm i}{\tilde q_{#1}}}}
\def\r#1{(\ref{#1})}
\nc{\qqq}[2]{e^{{\rm i}{\tilde q_{#1}}+{\rm i}{\tilde q_{#2}}}}
\nc{\sih}[2]{\sinh\frac{1}{2}(\g(#1)+\g(#2))}

\nc{\ra}{\rangle} \nc{\BAR}{\begin{array}} \nc{\EAR}{\end{array}}
\nc{\bdm}{\begin{displaymath}} \nc{\edm}{\end{displaymath}}
\nc{\be}{\begin{equation}} \nc{\ee}{\end{equation}}
\nc{\beq}{\begin{equation*}} \nc{\eeq}{\end{equation*}}
\nc{\ba}{\begin{array}} \nc{\ea}{\end{array}}
\nc{\bea}{\begin{eqnarray}} \nc{\eea}{\end{eqnarray}}

\nc{\hu}{{\bf u}}
\nc{\hh}{{\hat h}}\nc{\bl}{\boldsymbol{\lambda}}\nc{\hv}{{\bf v}}
\nc\si{{\mathrm{s}}}
\nc{\TS}{{\tilde S}}
\nc{\A}{\mathcal{A}} \nc{\Sc}{\mathcal{S}}  \nc{\N}{\mathcal{N}}
\nc{\rL}{{\rm L}}   \nc{\rR}{{\rm R}}
\nc{\fba}{{\zeta}}
\nc{\qs}{Q} \nc{\lms}{s}
\def\sx#1{\sigma^x_{#1}} \def\sz#1{\sigma^z_{#1}}
\def\qu{{\sf q}}
\def\pu{{\sf p}}
\def\SS{{S}}
\def\RRR{{\sf R}}
\nc\ve{{\varepsilon}}
\nc\mun{\nu}

\begin{document}
\title[Form-factors in the BBS model II: Ising model on the finite lattice]
{Form-factors in the Baxter--Bazhanov--Stroganov model II:
Ising model on the finite lattice}
\author{G von Gehlen$^\dag$,~ N Iorgov$^\ddag$,~ S~Pakuliak$^{\sharp\flat}$,  V~Shadura$^\ddag$
 and Yu~Tykhyy$^{\ddag\S}$}
\address{$^\dag$\ Physikalisches Institut der Universit\"at Bonn,
Nussallee 12, D-53115 Bonn, Germany}
\address{$^\ddag$ Bogolyubov Institute for Theoretical Physics, Kiev 03680,
Ukraine}
\address{$^\sharp$\ Bogoliubov Laboratory of Theoretical Physics,
Joint Institute for Nuclear Research, Dubna 141980, Moscow region,
Russia}
\address{$^\flat$\ Institute of Theoretical and Experimental Physics,
Moscow 117259, Russia}
\address{$^\S$\ Laboratoire de Math\'ematiques et Physique Th\'eorique CNRS/UMR 6083,
Universit\'e de Tours, Parc de Grandmont, 37200 Tours, France}
\ead{gehlen@th.physik.uni-bonn.de,
iorgov@bitp.kiev.ua, pakuliak@theor.jinr.ru,
shadura@bitp.kiev.ua, tykhyy@bitp.kiev.ua}
\bigskip
\begin{flushright}
\emph{Dedicated to Professor Anatoly Bugrij on the occasion of his 60-th birthday}
\end{flushright}

\begin{abstract}
We continue our investigation of the
Baxter--Bazhanov--Stroganov or $\tau^{(2)}$-model using the method of separation of variables
 \cite{gips,gipst1}. In this paper we derive for the first time
 the factorized formula for form-factors of the Ising model on a
 finite lattice conjectured previously by A.~Bugrij and O.~Lisovyy in
\cite{BL1,BL2}. We also find the matrix elements of the spin operator
for the finite quantum Ising chain in a transverse field.
\end{abstract}
\hspace*{2.5cm}{\small \today}\hspace*{12mm}   \submitto{\JPA}
\vspace*{-11mm} \pacs{75.10Hk, 75.10Jm, 05.50+q, 02.30Ik}

\section{Introduction}

The Baxter--Bazhanov--Stroganov (BBS) model \cite{BaxInv,BS}
(also called the $\tau^{(2)}$-model, see e.g. \cite{B_tau,Ro07})
is associated to the cyclic $L$-operators \cite{Kore,BS,Tara} which act in a two-dimensional auxiliary space
\be\label{Lop}
L_j(\lm)=\left( \ba{cc}
1+\lm\, \kp_j\, \hv_j &\hx  \lm\, \hu_j^{-1} (a_j-b_j\, \hv_j)\\
\hu_j\, (c_j-d_j\, \hv_j) &\hx \lm\, a_j\, c_j + \hv_j\, {b_j\, d_j}/{\kp_j }\ea \right)\,,
\ee
where $\lm$ is the spectral parameter, and at each site $j=1,\ldots,n$ we have
five parameters $a_j$, $b_j$, $c_j$, $d_j$ and $\kp_j$.
At each site there is also an ultra-local Weyl algebra with elements $\hu_j$ and
$\hv_j$, obeying
\[\fl \bu_j \bu_k=\bu_k \bu_j,\hx\! \bv_j\bv_k=\bv_k \bv_j,\hx\bu_j
\bv_k=\om^{\delta_{j,k}}\bv_k\bu_j,\hx\!\om=e^{2\pi i/N},\hx\!\bu_k^N=\bv_k^N=1,\hx\! N\in\mathbb{Z}_{\ge 2}\,.\]
Since $\om$ is a root of unity, the Weyl operators can be represented naturally by matrices acting
in the tensor product $\lk{\mathbb{C}^N}\rk^{\otimes n}$ \cite{gips,gipst1}.
The monodromy matrix of the model is defined by
\be\label{mm}  {T}_n(\lm)\,=\,L_1(\lm)\,L_2(\lm)\,\cdots\, L_n(\lm)=
\left(\begin{array}{cc} A_n(\lambda)&B_n(\lambda)\\C_n(\lambda)&D_n(\lambda)
 \end{array}\right),\ee
and the transfer matrix is its trace in the auxiliary space
$ {\bf t}_{n}(\lambda)\:=\:\mbox{tr}\:T_n(\lm)$ and gives
rise to a set of commuting non-local and
non-hermitian Hamiltonians of the model:
\[{\bf t}_n(\lm)=A_n(\lambda)+D_n(\lambda)={\bf
H}_0\,+{\bf H}_1\lm\,+\cdots\,+{\bf H}_{n-1}\lm^{n-1}\,+{\bf
H}_{n}\lm^{n}.\]
Commutativity follows from the intertwining of the $L_k(\lm)$
by the asymmetric 6-vertex model $R$-matrix at root of unity.

In the present paper our focus will be on the case $N=2$. As it has been shown in \cite{BIS}, in this
case the BBS-model is related to a generalized Ising model with plaquette Boltzmann weights
\be W(\sig_1,\sig_2,\sig_3,\sig_4)\;=\;a_0\lk 1\, +\, {\textstyle\sum_{1\le i\le j\le 4}}
          \,a_{ij}\,\sig_i\sig_j\,+\,a_4\,\sig_1\sig_2\sig_3\sig_4\rk, \label{FFIsing}\ee
subject to the free-fermion condition $\;a_4\,=\,a_{12}\,a_{34}\,-\,a_{13}\,a_{24}\,+\,a_{14}\,a_{23}.$

Generalizing Sklyanin's method of Separation of Variables (SoV) \cite{Skly1,Skly2,KarLeb2},
in \cite{gips,gipst1} we have worked out a method to find common eigenvectors of the
Hamiltonians ${\bf H}_{m}$. This proceeds in two steps:\footnote{We consider a fixed chain length $n$,
         mostly suppressing the subscript $n$,
         which had been written explicitly in our previous papers, so
         $\:r_k \equiv r_{n,k}\:$ and $\:\rho_n\equiv \rho_{n,k}\:$ of \cite{gips,gipst1}, etc.}
\begin{itemize}
\item Since $\left[ B(\lambda),B(\mu)\right]=0$, the off-diagonal element $B(\lm)$
of the monodromy matrix \r{mm} gives rise to an auxiliary set of commuting operators ${\bf h}_m$:
$B(\lm)\:=\: {\bf h}_1 \lm\, +\, {\bf h}_2 \lm^2
+\cdots + \,{\bf h}_{n}\lm^{n}$. We construct their common right eigenvectors $|\Psi_{\bdr}\,\rangle\:$
(the left eigenvectors $\:\langle\Psi_{\bdr}|\:$ are obtained analogously)
by an inductive procedure over the chain size $n$,
starting from the one site model.
The eigenvalues of $B(\lm)$ form a polynomial in $\lm$ of degree $n$, and from the intertwining
relations we can show that
\be \hspace*{-2cm} B(\lm)\:|\Psi_{\bdr}\,\ra\: =\:\lm\, r_{0}\,\om^{-\rho_{0}}\,
\prod_{k=1}^{n-1}\lk\lm\,+\,r_{k}\,\om^{-\rho_{k}}\rk|\Psi_{\bdr}\,\ra,\hq \bdr=(\rho_o,\rho_1,\ldots,\rho_{n-1}),
\label{bzero}\ee
where the amplitudes $(r_0,r_1,\ldots,r_{n-1})$ can be expressed in terms of the parameters of the model
$a_j,\ldots,\kp_j$, and we can use the phases $\bdr$, $\;\:\rho_k\in \mathbb{Z}_N$,
of the zeros
\be  \mun_k\;=\;-\,r_{k}\:\om^{-\rho_{k}},\hs k=1,\ldots,n-1,  \label{Bzero}\ee
of the eigenvalue polynomial for labelling the eigenvectors.
\item Having solved the auxiliary problem, after a Fourier transformation of $\rho_0$ to $\rho$ ($\rho$
labels the $\mathbb{Z}_N$-charge sectors), the eigenvalue problem of ${\bf t}(\lm)=A(\lm)+D(\lm)$ is reduced to the solution of
Baxter equations. Using that $A(\mun_k)$ and $D(\mun_k)$ are raising and lowering operators on the
auxiliary states $|\Psi_{\bdr}\ra$, we find that the periodic eigenstates $\:|\Phi_{\rho,{\bf E}}\rangle\,$ in
\[     {\bf t}(\lm)|\Phi_{\rho,{\bf E}}\rangle\;=
           \;t^{(\rho)}(\lm|{\bf E})\:|\Phi_{\rho,{\bf E}}\rangle, \hs {\bf E}=\{E_1,\ldots,E_{n-1}\} \]
with the eigenvalue polynomial
$\:t^{(\rho)}(\lm|{\bf E})\,=\,E_0+E_1\lm+\cdots+E_{n}\lm^{n}\:$
(where $E_0$ and $E_n$ are directly known, see \r{E0n}),
are obtained via the kernels ${\cal Q}^\eR$ in
\be  |\Phi_{\rho,{\bf E}}\rangle\;=\;\sum_{\rho_0,\bdr'}\om^{-\rho\cdot\rho_0}\:{\cal Q}^\eR(\bdr'|\rho,{\bf E})\;
             |\Psi_{\rho_0,\bdr'}\ra,\hx \bdr'\:=\:(\rho_1,\ldots,\rho_{n-1}).\label{rgl}\ee
The crucial fact is now (SoV) that after splitting off a known function $f(\bdr')$,
  the $n-1$-variable function $\,{\cal Q}^\eR(\bdr'|\rho,{\bf E})\,$
factorizes into single variable functions $\:Q_k^\eR(\rho_k)$ (the $\rho_k$ are the components of $\bdr'$, we often
skip the
charge index $\rho$ of the ${Q}^{\eR}_k$ etc.):
\[  {\cal Q}^\eR(\bdr'|\rho,{\bf E})\;=\;f(\bdr')\;\:{\textstyle\prod_{k=1}^{n-1}}\;Q_k^\eR(\rho_k), \]
and the $Q^\eR_k(\rho_k)$ are determined by the Baxter equations ($k=1,\ldots,n-1$):
\be\fl  t^{(\rho)}(\mun_k\,|{\bf E})\,Q_k^\eR(\rho_k)\,=\,\Delta_k^+(\mun_k)\,Q_k^\eR(\rho_k+1)\,
           +\Delta_k^-(\om\mun_k)\,Q_k^\eR(\rho_k-1).\label{Bax} \ee
The corresponding Baxter equations for the left periodic eigenvector read
\be \fl t^{(\rho)}(\mun_k\,|{\bf E})\,Q_k^\eL(\rho_k)\,=\,\om^{n-1}\Delta_k^-(\mun_k)\,Q_k^\eL(\rho_k+1)\,
           +\om^{1-n}\Delta_k^+(\om\mun_k)\,Q_k^\eL(\rho_k-1)\,.\label{BaxL} \ee
\end{itemize}
The functions $\Delta_k^\pm$ are defined by
\bea \fl\Delta_k^+(\lm)&=&(\om^\rho/\chi_k)\,(\lm/\om)^{1-n}\:\prod_{m=1}^{n-1}\,F_m(\lm/\om)\,,\hq
\Delta_k^-(\lm)\:=\:\chi_k\:(\lm/\om)^{n-1}\:F_n(\lm/\om)\,, \label{Delta}\\
\fl &&\hs\hs F_m(\lm)\;=\;\lk\, b_m\,+\om a_m \,\kp_m \lm\rk\,
\lk \,\la\, c_m\,+d_m/\kp_m\, \rk. \label{qdetn}\eea
We shall not need the expression for $\chi_k$, see (43) of \cite{gipst1}, since this will cancel in our final formulas.
The existence of a non-trivial solution to \r{Bax}, \r{BaxL} is provided by a set of functional equations,
which determines the still unknown values ${\bf E}$.

In \cite{gipst1} we calculated the action of $\,\bu_n$, the Weyl operator at site $\,n\,$,
on an eigenvector $|\Psi_{\bdr}\ra$ of $B(\la)$ to have the form
\be \fl \qquad \bu_n|\Psi_{\bdr}\ra=g\:|\Psi_{\bdr}\ra+
{\textstyle \sum_{k=0}^{n-1}}\,g_k\:|\Psi_{\bdr^{+k}}\ra
\hx\mbox{with} \hx \bdr^{+k}\!=(\rho_0,\ldots,\rho_k+1,\ldots,\rho_{n-1}),\label{ushift}\ee
and $g$ and $g_k$ are certain functions depending on the
parameters $a_j,b_j,c_j,d_j,\kp_j$ and the components of $\bdr$.
Since in \cite{gips,gipst1} we also found a factorized expression
for the norm $\langle\Psi_{\bdr}|\Psi_{\bdr'}\ra$, we have the framework for calculating normalized matrix elements
$\langle\Psi_{\bdr}|\bu_n|\Psi_{\bdr'}\ra /\langle\Psi_{\bnul}|\,\Psi_{\bnul}\ra$,
where $\bnul=(0,0,\ldots,0)$. For
calculating matrix elements between periodic states $\langle\Phi_{\rho}|\bu_n|\Phi_{\rho'}\ra $, in addition we
also need the solutions $Q_k^{(\rho)}(\rho_k)$ of the Baxter equations.
These are available for $N=2$, and our main goal
is to obtain such periodic matrix elements in a factorized form. We achieve this by explicitly
performing the sums over the intermediate ${\mathbb Z}_2$ variables.

This paper is organized as follows: In the following Section 2 we recall the $N=2$ spin matrix element calculated
in \cite{gipst1} and
transform it into a much more compact form by performing the summation over an intermediate variable,
still keeping the model general and inhomogeneous. In order to proceed beyond this result,
in Section 3 we specialize the parameters of the BBS-model such
that we get the homogenous Ising model. We discuss the structure of the eigenvalues and the solutions
to the Baxter equations. The vanishing
of some transfer matrix eigenvalues at the zeros of $B(\la)$ requires us to distinguish four cases when solving
the Baxter equations.
Then in Section 4 we continue the evaluation of the matrix elements of the spin operator, until we are finally
able to perform the
multiple sum over intermediate spins.
 The derivation of the basic formula for the multiple spin summation is delegated to the Appendix.
Section 5 deals with the calculation of squares of the matrix elements. In Section 6 we include a special
case excluded in the
earlier Sections and give the final formulas for the matrix
element of spin operators in terms of the zeros of the transfer matrix and excited quasi-momenta.
Then we are ready to compare our result in Section 7 to a conjectured formula
of A.~Bugrij and O.~Lisovyy \cite{BL1,BL2}.
In Section 8 we apply the formulas of Section 6 to obtain
the matrix element of $\sigma^z$ for the finite quantum Ising chain in a transverse magnetic field.
In Section 9 we give our conclusions.

\section{Spin operator matrix element for the $N=2$ inhomogenous BBS model}\label{Inhom}

In \cite{gipst1} we derived a formula for the normalized matrix element of the spin operator between arbitrary states of the
periodic inhomogenous $N=2$ BBS-model.
For $N=2$ there are two $\mathbb{Z}_2$-charge sectors $\rho=0,1$ in \r{rgl}. Since $\bu_n$ is anticommuting with the charge
operator $\bV=\bv_1\bv_2\ldots\bv_n$, its only non-vanishing matrix elements are between periodic
states $|\Phi_\rho\,\rangle$ with different charge $\rho$. Let $Q^{{\rm L}(\rho)}$ and $Q^{{\rm R}(\rho)}$
be solutions to the Baxter equations \r{Bax}, \r{BaxL} and $r_k$ the zeros of the operator polynomial $B(\la)$,
see \r{bzero}.
Then our result in (58) of \cite{gipst1} can be written:
\be\label{mat-el}\fl
\frac{\langle\, \Phi_0\,|\bu_n|\,\Phi_1\,\rangle}
{\langle \,\tilde\Psi_{0,\bnul}\,|\,\tilde\Psi_{0,\bnul}\,\rangle}
\,=\,\sum_{\bdr'} \CN(\bdr')
\left(R_0(\bdr') \left(\frac{a_n}{\tilde r}(-1)^{\tilde  \rho'}-
\frac{\kp_1\kp_2\cdots\kp_{n-1} b_n}{r_{0}}\right) \!
+\sum_{k=1}^{n-1} R_k(\bdr') \right),
\ee
\be\label{NoN}\fl
\CN(\bdr')= (-1)^{n\tilde \rho'}\, \prod_{l<m}^{n-1}\:
\frac{r_{l}+r_{m}}{(-1)^{\rho_{l}}\, r_{l}\,+ (-1)^{\rho_{m}}\,r_{m}},
\quad\hq
R_0(\bdr')=\prod_{l=1}^{n-1}{\qs}^{{\rm L}(0)}_l(\rho_{l}){\qs}^{{\rm R}(1)}_l(\rho_{l}),
\ee
\bea \fl R_k(\bdr')&=&-\frac{a_n b_n c_n}{r_{0}}\;
Q^{\rL(0)}_k(\rho_{k}+1)\:Q^{\rR(1)}_k(\rho_{k})\prod_{l\ne k}^{n-1} Q^{\rL(0)}_l(\rho_{l})\:Q^{\rR(1)}_l(\rho_{l})\times
\ny\\ \fl && \times\;
\lk 1-\frac{d_n}{\kp_n c_n\, \mun_{k}}\rk
\frac{\mun_k^{n-1}\chi_k }{\prod_{s\ne k} (\mun_{k}-\mun_{s})}\,,\label{Rk}
\eea
with $\mun_k=-r_k(-1)^{\rho_k}$, $\tilde r=r_{0}\,r_{1}\cdots r_{n-1}$ and
$\tilde\rho'=\sum_{k=1}^{n-1}\:\rho_k$.
The different terms in \r{mat-el} have the following origin:
$\CN(\bdr')$ is a normalization factor due to the convenient (since it avoids further factors) choice of normalizing by
$\langle \tilde\Psi_{0,{\bf 0}}|\tilde\Psi_{0,{\bf 0}}\rangle$ of the auxiliary system.
Here $|\tilde\Psi_{\rho,\bdr'_n}\rangle=|\Psi_{0,\bdr'_n}\rangle+(-1)^\rho|\Psi_{1,\bdr'_n}\rangle$,
where $|\Psi_{\rho_0,\bdr'_n}\rangle$ is an eigenvector defined in \r{bzero},  see also (35) in \cite{gipst1}.
For the terms in \r{mat-el} recall \r{ushift}. The terms at $R_0(\bdr')$
correspond to $g$ and $g_0$ in \r{ushift}. The sum over $k$ and the expression for $R_k(\bdr')$
arise from the shift in the index $\rho_k$ and the coefficients $g_k$.

In the remaining part of this Section we now show that the sum over $k$ can be performed, leading to the much simpler expression \r{mat-res}, \r{Rs}.

We start rewriting the factors of the $R_k(\bdr')$:
\bea \fl \lefteqn{-\frac{a_n b_n c_n}{r_{0}}\;\mun_k^{n-1}\chi_k
   \lk 1-\frac{d_n}{\kp_n c_n\, \mun_{k}}\rk\;=\;
   \frac{(-1)^{n-1} b_n}{r_0\,\kp_n\,\mun_k\,(\mun_k\,+\fba_n)}\;{\Delta_k^\nul}^-(\mun_k)}\ny\\ \fl &&
=\frac{(-1)^{n-1}  b_n}{r_0\,\kp_n\,\mun_k\,(\mun_k\,+\fba_n)}\;
\lb\frac12\lk\Delta_k^{(0)+}(-\mun_k)+\Delta_k^{(0)-}(\mun_k)\rk+
 \frac12\lk\Delta_k^{(1)+}(-\mun_k)+\Delta_k^{(1)-}(\mun_k)\rk\rb\ny\\
 \fl &&
 =\frac{(-1)^{n-1} b_n}{2\,r_0\,\kp_n\,\mun_k\,(\mun_k\,+\fba_n)}\,\lk
   \frac{(-1)^{n-1}t^\nul(\mun_k)\:\:Q_k^{\rL(0)}(\rho_k)}{Q_k^{\rL(0)}(\rho_k+1)}\;
  +\:\frac{t^\one(-\mun_k)\:\:Q_k^{\rR(1)}(\rho_k+1)}{Q_k^{\rR(1)}(\rho_k)}\rk\!\!,
\label{simpli}\eea
where we define $\fba_k=b_k/(a_k\kp_k)$ and use $\Delta_k^{(\rho)\pm}(\lm)$
from \r{Delta}, pointing out the explicit dependence on $\rho$.
For obtaining the first two lines of \r{simpli} we used
\[\fl \Delta_k^{(0)-}\!(\mun_k)=\,\Delta_k^{(1)-}\!(\mun_k)=\,
    \chi_k(-\mun_k)^{n-1}\,(b_n\,+\,a_n\,\kp_n\,\mun_k)\,(-\mun_kc_n\,+\,d_n/\kp_n),   \]
and $\Delta_k^{(0)+}(-\mun_k)=-\Delta_k^{(1)+}(-\mun_k)$.
To get the third line of \r{simpli} we used the Baxter equations \r{Bax}, \r{BaxL},
 where for $N=2$ we have $\om=-1$ and $\rho_k+1=\rho_k-1$ $\mbox{mod}\ \mathbb{Z}_2$:
\[\fl
\frac {\qs^{{\rm L}(0)}_k(\rho_{k})}{\qs^{{\rm L}(0)}_k(\rho_{k}+1)}=
\frac{\Delta_k^{(0)+}(-\mun_k)+\Delta_k^{(0)-}(\mun_k)}{(-1)^{n-1} t^{(0)}(\mun_k)}\,,\quad
\frac {\qs^{{\rm R}(1)}_k(\rho_{k}+1)}{\qs^{{\rm R}(1)}_k(\rho_{k})}=
\frac{\Delta_k^{(1)+}(-\mun_k)+\Delta_k^{(1)-}(\mun_k)}{ t^{(1)}(-\mun_k)}\,.
\]
Then, using \r{simpli}, \r{Rk} becomes:
\bea \fl \lefteqn{R_k(\bdr')\;=\;\frac{b_n}{2\,r_0\,\kp_n}\;\:
    \frac{1}{\mun_k\,(\mun_k\,+\fba_n)\,\prod_{s\neq k}(\mun_k\,-\mun_s)}\;\lk
     t^\nul(\mun_k)\:R_0(\bdr')+
 \phantom{\prod_{l\neq k}^{n-1}}\right.} \ny\\ \hspace*{-17mm} +\;\left.
   (-1)^{n-1}t^\one(-\mun_k)\:\:Q_k^{\rL(0)}(\rho_k+1)\,Q_k^{\rR(1)}(\rho_k+1)\:
   \prod_{l\neq k}^{n-1}Q_l^{\rL(0)}(\rho_l)\:Q_l^{\rR(1)}(\rho_l)
\rk\!.\label{simpp}
\eea
Now the sum over $k$ in \r{mat-el} can be performed
using an identity (c.f. the Appendix of \cite{gipst1}), valid for any polynomial $f(x)$
of degree less than $n+1$ and for any $n+1$ non-coincident points $x_k$:
Consider a polynomial $f(x)=f_n\,x^n+\ldots+f_0$, its interpolation through
the points $x_1,\ldots,x_{n+1}$, and focussing attention on the coefficient of $x^n$:
\be    f(x)\;=\;\sum_{k=1}^{n+1}\;f(x_k)\;\prod_{s\neq k}^{n+1}\;
                    \frac{x-x_s}{x_k-x_s}\,,\qquad
 f_n\;=\;\sum_{k=1}^{n+1}\;\frac{f(x_k)}{\prod_{s\neq k}^{n+1}(x_k-x_s)}\,.
 \label{sumru}\ee
For calculating the sum over $k$ of the first term in the parentheses of \r{simpp}, in \r{sumru}
we take $f(x)=t^\nul(x)$ and
$(x_1,\ldots$, $x_{n-1}$, $x_n$, $x_{n+1})=(\mun_1,\ldots$, $\mun_{n-1},0,-\fba_n)$. Thus we get
\be \fl \sum_{k=1}^{n-1}\:\frac{t^\nul(\mun_k)}
          {\mun_k\,(\mun_k\,+\fba_n)\,\prod_{s\neq k}(\mun_k\,-\mun_s)}\,=\,
 E_n^\nul -\frac{t^\nul(0)}{\fba_n\prod_{s=1}^{n-1}\,(-\mun_s)}
           \,-\frac{t^\nul(-\fba_n)}{-\fba_n\prod_{s=1}^{n-1}(-\fba_n\,-\mun_s)},\label{Rk0}
\ee
where $\:E_n^\nul\:$ is the leading coefficient of $\;t^\nul(\la)$.

For the second term of the last line of \r{simpp}
we will not perform the summation over $k$ directly. Instead,
for each $\bdr'$ for which we make the summation of the first term, we
take for the summation over $k$ the second term of \r{simpp}
corresponding to ${\bdr'}^{+k}$ (which entails $\mun_k\rightarrow-\mun_k$) in \r{mat-el}.
Collecting all such terms and taking into account the changes which come from
$\CN({\bdr'}^{+k})/\CN({\bdr'})$ we perform the summation over $k$.
Together with \r{Rk0}, we get
\bea \fl \lefteqn{\frac{b_n}{2\,r_0\,\kp_n}\:R_0(\bdr')\:
    \lk E_n^\nul\;-\:\frac{t^\nul(0)}{\fba_n\,
  \prod_{s=1}^{n-1}\:(-\mun_s)}\;+\;\frac{t^\nul(-\fba_n)}
    {\fba_n\,\prod_{s=1}^{n-1}(-\fba_n\,-\mun_s)}\:+\right.}\ny\\  &&\left.\hs\hs\hq
-\;E_n^\one\;-\:\frac{t^\one(0)}{\fba_n\,
  \prod_{s=1}^{n-1}\:(-\mun_s)}\;+\;\frac{t^\one(\fba_n)}{\fba_n\,\prod_{s=1}^{n-1}(\fba_n\,-\mun_s)}\rk\!,
  \label{EEn}\eea
The leading and the constant coefficients of $t^{(\rho)}(\la)$ can be read off directly from \r{Lop},\r{mm}:
\be \fl E_n^{(\rho)}=\prod_{l=1}^n\,a_l\,c_l+(-1)^\rho\:\prod_{l=1}^n\:\kp_l,\hs
 E_0^{(\rho)}=t^{(\rho)}(0)=1+(-1)^\rho\prod_{l=1}^n\,(b_l d_l/\kp_l). \label{E0n}\ee

Inserting these results into \r{mat-el} we find that the $E^{(\rho)}_n$ and $E^{(\rho)}_0$ terms in
\r{EEn} just cancel the terms of the bracket at $R_0(\bdr')$ in \r{mat-el}, and we get simply
\be \frac{\langle\,\Phi_0\,|\,\bu_n\,|\,\Phi_1\,\rangle}
{\langle\, \tilde\Psi_{0,{\bf 0}}\,|\,\tilde\Psi_{0,{\bf 0}}\,\rangle}
\;=\;\frac{a_n}{2\,r_0}\sum_{\bdr'\in{\mathbb Z}_2^{n-1}} \CN(\bdr')\,R_0(\bdr')\:R(\bdr')\label{mat-res}\ee
with\\[-11mm]
\be R(\bdr')\;=\;\frac{t^\nul(-\fba_n)}
    {\prod_{l=1}^{n-1}(-\fba_n+(-1)^{\rho_l}r_l)}\:+\:
\frac{t^\one(\fba_n)}{\prod_{l=1}^{n-1}(\fba_n+(-1)^{\rho_l}r_l)}.\label{Rs}\ee
We shall write $\;\;R(\bdr')\:=\:R^\nul(\bdr')\:+\:R^\one(\bdr')\;$ when we have to refer to the separate terms
on the right hand side of \r{Rs}.
Despite the simple appearance, for the general inhomogenous $N=2$ BBS-model,
performing the multiple sums over the ${\mathbb Z}_2$ variables seems to be a presently impossible task.
However, restricting ourselves to the homogenous model with the parameters satisfying
\be\fl a_j\:=\:c_j\;=\;a,\hs b_j\:=\:d_j\;=\;b,\hs \kp_j\,=\,1
 \hq\mbox{for}\hx j\,=1,\ldots,n-1.\label{Isi} \ee
we are able to evaluate \r{mat-res} with \r{NoN}, \r{Rs} completely, as will be shown in Section \ref{Calc}.

\section{Homogeneous Ising model}

In all following Sections we consider only the $N=2$ case of the model defined by \r{Lop} with \r{Isi}.
For a fixed chain length $n$,
we are left with only the two parameters $a,\;b$ and the spectral parameter $\la$.
 For $N=2$ we have $\om\,=-1$ and we represent the Weyl
operators $\bu_k,\;\bv_k$ by Pauli matrices acting at the $k$-th site. So, now our model is defined by\\[-4mm]
\be\label{L-Ising}
L_k(\lm)=\lk\ba{cc} 1\,+\lm\,\sx{k} &  \lm \,\sz{k}\, (a\,-b\, \sx{k})\\[2mm]
\sz{k}\, (a\,-b\, \sx{k}) & \lm a^2 \,+ \sx{k}\, b^2\ea \rk\,.\ee
Fixing the spectral parameter to the value $\lm=b/a$, the $L$-operator \r{L-Ising} degenerates
\[  L_k(b/a)\;=\;(1\,+\, \sx{k}\:b/a)\lk \ba{c} 1\\ a\, \sz{k} \ea \rk\lk \ba{cc} 1\,,& b\,\sz{k}\ea \rk \]
and the transfer matrix can be put into the standard Ising form
\bea\fl\label{Isitra}\lefteqn{
{\bf t}(b/a)=\tr\, L_1(b/a)\,L_2(b/a)\cdots L_n(b/a)\:=\:
\prod_{k=1}^n (1+\sx{k} \cdot {b}/{a})\cdot \prod_{k=1}^n (1+\sz{k-1}\sz{k}\cdot a\,b)}\ny\\
    \fl \hspace*{6.2cm}&\sim&\:\exp{\lk {\textstyle\sum_{k=1}^n}\, K^*_x\, \sigma^x_k\rk}\:
\exp{ \lk{\textstyle\sum_{k=1}^n}\, K_x \,\sigma^z_{k-1}\,\sigma^z_k\rk}\,,
\eea
if we use periodic boundary conditions $\sz{n+k}\equiv\sz{k}$ and identify
\be
e^{-2K_y}\,= \tanh K^*_x\,=\,{b}/{a}\,,\quad\; \tanh K_x\,=\,a\!\,b\,.
\label{KKIsing}\ee
So at $\la\,=\,b/a\;$ we call the model \r{L-Ising} the Ising model. If we don't fix
the spectral parameter to this special value, we shall talk of the ``generalized Ising model''. However,
transfer matrix eigenstates are independent of the choice of $\lm$.

\subsection{Structure of the eigenvalues}

In \cite{gips} the eigenvalues of the transfer matrix ${\bf t}(\lm)=\tr\:L_1(\lm)\,\cdots\,L_n(\lm)$
 with $L_k(\lm)$ given by \r{Lop} for $N=2$ and homogeneous parameters, have been calculated
from the functional relations. From $\mathbb{Z}_2$-invariance $\;{\bf t}(\lm)\:$ commutes
with the $\mathbb{Z}_2$-charge operator ${\bf V}\,=\,\sx{1}\,\sx{2}\,\ldots\,\sx{n}\,.$ Since ${\bf V}^2=1$, the
space of eigenstates of $\;{\bf t}(\lm)\:$ decomposes into two sectors according to the
eigenvalues $(-1)^\rho$ (where $\rho=0,\:1$) of ${\bf V}$.
The sector $\rho=0$ is called the NS-sector, $\rho=1$ the R-sector. The $2^n$ eigenvalues
can be written (we specialize assuming \r{Isi}):
\be\label{tmgen}\fl
t^{(\rho)}(\lm)=(a^{2n}\!+(-1)^\rho)\,\prod_{\qu}\,(\lm\,+(-1)^{\sig_\qu}\,\lms_{\qu}),\hq
\lms_{\qu}\,=\,\lms_{-\qu}\,=\,\sqrt{\frac{b^4\,-2\, b^2\cos\qu+1}{a^4\,-2\, a^2\cos\qu+1}}\,,
\ee
where the quasi-momentum $\qu$ in each sector takes $n$ values
$\:\qu\,=\,\frac{2\pi}{n}\,m\:$ with
$m$ integer (half-integer) for the R (NS)-sectors.
If $\sig_\qu=0$ the quasi-momentum $\qu$
is called unexcited, for $\sig_\qu=1$ it is called excited.
In the NS (R) sector, the eigenstates of ${\bf t}(\lm)$ have an even (odd) number of
excitations: $\prod_{\qu}\,(-1)^{\sig_\qu}\,=\,(-1)^\rho$.

For $\qu=0$ (this occurs only for the R-sector) and for $\qu=\pi$  we define
\be  \lms_{0}=\frac{b^2-1}{a^2-1}\,,\quad \hs\lms_{\pi}=\frac{b^2+1}{a^2+1}\,.\label{s}\ee
The quasi-momentum $\qu=\pi$ is in the R sector for $n$ even, it is in the NS sector for $n$ odd.
The different presence of factors $(\lm\,\pm \lms_0)$ and $(\lm\,\pm\lms_\pi)$ in \r{tmgen} for $n$
even or odd often makes it necessary to
consider the cases of even $n$ and odd $n$ separately.

Sometimes we shall use the notation $\lm_{\qu}:=(-1)^{\sigma_{\qu}}\lms_{\qu}$.

\subsection{State vectors from Baxter equations}\label{zweizwei}

In order to obtain the eigenvectors of the transfer matrix ${\bf t}(\lm)$,
we have to solve Baxter's equations \r{Bax} and \r{BaxL}.
As input we use the corresponding eigenvalues $t^{\rho}(\lm)$
which are specified by the values $\sig_\qu$ for all $\:\qu\:$ in the sector $\rho$,
see \r{tmgen}. Solving Baxter's equations, we should use the values $t^{\rho}(\pm r_k)$
of these polynomials at the values
 $\pm r_{k}$ of the roots of the eigenvalue polynomials of the operator
$B_n(\lm)$ (which is the off-diagonal element of the monodromy matrix) given by the formula
 (A7) of \cite{gips}. For our special parameters \r{Isi} and $N=2$ the $r_k$
are simply related with the $\lms_{\qu}$:
\be r_k\;=\;\lms_{q_k}\,,\hs q_k\,=\,\pi k/n,
\hs k=1,\ldots,n-1.\label{rzero}\ee
This means that for our special choice of parameters \r{Isi},
the zeros of $t^{\rho}(\lm)$ may coincide
with the $r_k$, giving rise to the vanishing of the left hand sides of \r{Bax} and \r{BaxL}.
At the parameters \r{Isi} all $\:F_m$ are equal: $F_m(\lm)\,=\,F(\lm)$ and from \r{qdetn} we obtain
\be\label{shchi}
F(\la)\:=\:b^2\,-a^2\,\la^2,\hs\chi_k^2\: r_k^{2(n-1)}=(-1)^{n+k+1}F^{n-2}(r_k)\,.\ee

Let us compare two sets: the set $\{q_k\}$ which parameterizes the roots $r_k$, and
the set of all possible quasi-momenta $\{\qu\}$.
The latter set is divided into two sub-sets: the NS and the R sectors.
The NS-sector contains pairs of quasi-momenta $\{q_k,-q_k\}$ for odd $k$ and the R-sector includes
the pairs $\{q_k,-q_k\}$ for even $k$. The quasi-momentum $\qu=0$ always belongs to the R-sector,
and $\qu=\pi$ belongs to the R-sector for even $n$ and to the NS-sector for odd $n$.

The solutions of Baxter's equations for the case of the Ising model were found
in \cite{gipst1}. Here we recall the final result.
For a fixed sector $\rho$ and the eigenvalue polynomial $t^{\rho}(\lm)$
we have to solve $n-1$ systems of Baxter's equations \r{Bax} (or \r{BaxL})
numerated by the integers $k=1,\ldots,n-1$. With respect to these data we have to distinguish
four cases:\\[2mm]
\noindent
\underline{$(-1)^\rho=(-1)^{k}$:}\\[1mm]
(i)\hq $\,t^{\rho}(r_k)\ne 0$ and $t^{\rho}(-r_k)\ne 0$:
\[Q_k^{\rm L,R}(0)\,=\,1\,,\hq
Q_k^{\rm L,R}(1)\:=\:\frac{(-1)^{n-1}t^{\rho}(-r_k)}{2 \chi_k\:r_k^{n-1}\:F(r_k)}\,.
\]
\vspace*{2mm}
The other three cases occur for\\
\underline{$(-1)^\rho=(-1)^{k-1}:$}$\hs$
\\[1mm]
(ii)\hq
$t^{\rho}(r_k)\ne 0,\ t^{\rho}(-r_k)=0$:
$t^{\rho}(\lm)$ contains a factor $(\lm+r_k)^2$
(both $\qu=\pm q_k$ not excited), we may normalize
\[Q^{\rm L,R}_k(0)=1, \hq Q^{\rm L,R}_k(1)\:=\:0\,.\]
(iii)\hq $t^{\rho}(r_k)=0,\ t^{\rho}(-r_k)\ne 0$:
$t^{\rho}(\lm)$ contains a factor $(\lm-r_k)^2$ (both $\qu=\pm q_k$ are excited),
       we cannot choose $Q^{\rm L,R}_k(0)\,=\,1$, but we may normalize
       \[Q^{\rm L,R}_k(0)\:=\:0\,,\hq Q^{\rm L,R}_k(1)\:=\:1\,.\]
(iv)\hq
$t^{\rho}(r_k)=t^{\rho}(-r_k)=0$: $t^{\rho}(\lm)$
contains $(\lm^2-r_k^2)$
(either $\qu=+q_k$ or $\qu=-q_k$ is excited): A L'H{\^o}pital procedure as described
in \cite{gipst1} is required (in order to obtain eigenvectors
of the translation operator), leading to
\[\fl
Q^{\rm R}_k(0)=Q^{\rm L}_k(0)=1\,,\hx
Q^{\rm R}_k(1)=-Q^{\rm L}_k(1)=
\frac{(-1)^{n+\sig_{q_k}+1}  2{\rm i}\sin{q_k}\,{t}^{\rho}_{\check q_k}(-r_k)}
{ n\, \chi_k\:r_k^{n-1}\: A(q_k)}\,,\]
where
\be \fl
t^{\rho}(\lm)\:=\:{ t}^{\rho}_{\check q_k}(\lm)\;
(\lm+(-1)^{\sig_{q_k}}\lms_{q_k})(\lm-(-1)^{\sig_{q_k}}\lms_{-q_k}),
\hq A(\qu)\:=\:a^4-2a^2\cos{\qu}+1. \label{tcheck}\ee

In the next Sections we shall restrict ourselves to calculate only transitions between
eigenvectors allowing the normalization $Q^{\rm L,R}_k(0)=1,$ postponing to Section 6 the consideration of
eigenvectors for which $t^{\rho}(\lm)$ contains factors $(\lm-r_k)^2$, i.e. the eigenvectors involving case
(iii) above.

As already observed at the beginning of Section \ref{Inhom}, the non-vanishing spin matrix elements have left and
right eigenstates from different sectors. Let
$t^{(0)}$ and $t^{(1)}$ be the corresponding eigenvalue-polynomials.
With respect to these two polynomials
we define\\
$\hspace*{27mm} k\in\CDC^{(\rho)}$ if $t^{\rho}$ has a factor $(\lm+r_k)^2$, i.e. we have case (ii),\\
$\hspace*{27mm} k\in\WD^{(\rho)}\;\;$ if $t^{\rho}$ has a factor $(\lm-r_k)^2$, case (iii), and\\
$\hspace*{27mm} k\in\CD^{(\rho)}\;\;$ if $t^{\rho}$ has a factor $(\lm^2-r_k^2)$, i.e. we have case (iv).\\
By $D=|\CD|$ we denote the number of elements in $\CD=\CD^{(0)}\cup \CD^{(1)}$, similarly for $\CDC$, etc.

\section{Calculation of the matrix element of $\sig_n^z$ in the homogeneous Ising model}\label{Calc}

We now start to evaluate \r{mat-res} with \r{NoN} and \r{Rs} in our simplified model where
\be \fba\:=\:b/a,\hs r^2_{0}\,=\;(a^2-b^2)(a^{4n}-1)/(a^4-1).\label{avier}\ee
 We have to observe that in the derivation of \r{mat-el}
given in \cite{gipst1}, generic BBS-parameters leading to $t^\rho(r_k)\neq 0$ were assumed,
and the solutions to the Baxter equation
were normalized to $Q^{\rm L,R}_k(0)=1$. As we have seen in Section~3.2,
in the case (iii) this normalization is not possible
for the special parameters \r{Isi}. In order not to complicate the derivation,
in the following part of this Section we
shall simply exclude state vectors containing $k\in\WD$, adding the changes necessary
for $k\in\WD$ in Section \ref{Final}.
Also in this section we shall omit the superscripts L and R
in the notations of $Q^{{\rm L,R}(\rho)}_k(\rho_k)$
supposing that the left/right eigenvectors are from NS/R-sectors as they appear
in \r{mat-res}.

Consider $R_0(\bdr')$. Always one of the factors in
    $Q_l^{(0)}(\rho_l)Q_l^{(1)}(\rho_l)$ is from case (i) above, and excluding $l\in\WD$,
the other is from either (ii) or (iv). So always $Q_l^{(0)}(0)Q_l^{(1)}(0)\,=\,1$.
For $l\in \CDC$ we have $\qs^{(0)}_l(1){\qs}^{(1)}_l(1)=0$ since either
$\qs^{(0)}_l(1)=0$ or ${\qs}^{(1)}_l(1)=0$ depending on the parity of $l$.
So, in (\ref{mat-res}) the summation reduces to the summation over $\rho_l$ for $l\in \CD$, with
$\rho_l=0$ fixed for $l\in \CDC$.

\subsection{Calculation of $\;R(\bdr')$}\label{calcR}

Let us show that a common factor can be extracted from the two terms of \r{Rs}.
We first consider the case of odd $n$ where $\qu=0$ appears in the R-sector and $\qu=\pi$ in the NS-sector.

We start with the first term $R^\nul(\bdr')$ in \r{Rs}.
Now from \r{tmgen} the NS eigenvalue polynomial $t^{(0)}(\lm)$ for odd $n$ is
\be \fl \mbox{NS},\;\;n\;\mbox{odd:}\hx t^{(0)}(\lm)\:=\:(a^{2n}+1) (\lm+(-1)^{\sigma_\pi}s_\pi)
\prod_{k \in \CDC^{(0)}}
               (\lm+r_{k})^2 \prod_{l\in\CD^{(0)}} (\lm^2-r_{l}^2)\,,\label{tNS}\ee
(for even $n$ omit the bracket with $s_\pi$) since in the NS-sector
only odd $k$ appear, and these fall into one of the classes (ii) and (iv), class (iii) being
momentarily excluded.
We insert $t^\nul(-\fba)$ from \r{tNS} and
decompose the denominator product over $l$ in its even-$l$ and odd-$l$ parts. We write the odd part
as $l\in\CD^\nul\cup\CDC^\nul$ since for $\rho=0$ in cases (ii) and (iv) $l$ must be odd (recall,
we still exclude case (iii)):
\\[-2mm]
\bea\fl \lefteqn{R^{(0)}(\bdr')\,=\,(a^{2n}+1)\,\frac{(-\fba+(-1)^{\sig_\pi}s_\pi)}
         {\prod_{l\,{\rm even}}(-\fba+(-1)^{\rho_l}r_l)}\;
 \frac{ \prod_{k\in\CDC^\nul}(-\fba+r_k)^2\:\prod_{l\in\CD^\nul}(\fba^2-r_l^2)}
 {\prod_{k\in\CDC^\nul}(-\fba+r_k)\:\prod_{l\in\CD^\nul}(-\fba+(-1)^{\rho_l} r_l)}}
 \ny\\[3mm] \fl&&\hspace*{6cm}\times\;
         \frac{\prod_{m\in\CDC^\one}(\fba+r_m)\prod_{m\in\CD^\one}(\fba+(-1)^{\rho_m}r_m)}{\prod_{m\,{\rm even}}
        (\fba+(-1)^{\rho_m}r_m)}\,.
\label{Rodd}\eea
In the last line we put a factor unity, written as quotient of upstairs a product over $m\in\CD^\one\cup\CDC^\one$
 and downstairs over $m\,{\rm even}$. In the $\CDC$ terms we omitted the factor
 $(-1)^{\rho_k}$  since from (ii)
   this contributes only if $\rho_k=0$. Now several cancellations take place, resulting in
\be \fl R^{(0)}(\bdr')\:=\,
\frac{(a^{2n}+1)\; (-\fba+(-1)^{\sigma_\pi}s_\pi)}
{(-1)^{|{\CD}^{(0)}|}\prod_{{l\:{\rm even}}}
(-\fba^2\,+\,r_l^2)}\;\prod_{k\in \CDC}  ((-1)^k \fba+r_k)\;
                 \prod_{l\in {\CD}}(\fba\,+(-1)^{\rho_l}r_l)\,.\label{Rnn}\ee
Observe that now the $\bdr'$-dependence appears only in the last product over $\CD$.
This happens because all $l$-odd terms cancel and because $\CDC^\nul$ allows only $\rho_l=0$.
In the denominator we use $\fba=b/a$, \r{tmgen}, \r{rzero} and
$\prod_{l\;{\rm even}}A(q_l)\:=\:(a^{2n}-1)/(a^2-1)$
to obtain
\be \textstyle\prod_{l\:{\rm even}}(-\fba^2\,+\,r_l^2)=\lk(\fba^2-1)(a^2b^2-1)\rk^{(n-1)/2}(a^2-1)/(a^{2n}\:-\:1)\,.
\label{sumA}\ee

The second term in \r{Rs} can be evaluated analogously. We insert $t^{(1)}(\fba)$ from
\be
\fl\mbox{R, $\;\;n$ odd:}\hq
t^{(1)}(\lm)\:=\:(a^{2n}-1) (\lm\,+(-1)^{\sigma_0}s_0)\!\!\prod_{k \in \CDC^{(1)}} (\lm\,+r_{k})^2
\prod_{l\in \CD^{(1)}} (\lm^2\,-r_{l}^2)
\label{tR}\ee
and use $\;\:\prod_{l\;{\rm odd}}A(q_l)\:=\:(a^{2n}+1)/(a^2+1)\;$ to get finally for $n$ odd:
\bea\fl R^{(n\,{\rm odd})}(\bdr')&=&\lk (-1)^{\sig_\pi} (a^2+1)
\lk -\fba\,\,+(-1)^{\sig_\pi}\lms_\pi\rk\;
{\textstyle \prod_{l\in \CD}}((-1)^{\rho_l}\,r_l\,+\fba)\right. \ny\\
\label{rRo}\fl
&&\left.-(-1)^{\sig_0} (a^2-1) \lk \fba\,+(-1)^{\sig_0}\lms_0\rk\;
{\textstyle \prod_{l\in \CD}}((-1)^{\rho_l}\,r_l\,-\fba)\rk\;\RRR
\eea
with (using also \r{avier})
\be\label{RRR}
\RRR=r_0^2 \aab^{-1} (\aab\abb)^{-(n-1)/2}  a^{n-1}{\textstyle \prod_{k\in \CDC}\; ((-1)^k\,\fba\,+r_k)}\,,
\label{oddR}\ee
where $\aab\:=\:a^2-b^2$, $\abb\:=\:1-a^2b^2$, and
$(-1)^{|\CD^{(0)}|}=(-1)^{\sig_\pi}$, $(-1)^{|\CD^{(1)}|}=-(-1)^{\sig_0}$ in the case of odd $n$.

The case of $n$ even is less symmetric between $R^\nul$ and $R^\one$ since
now both $\qu=0\,$ and $\qu=\pi\,$ are in the R sector,
none of them in NS.~
So the term containing $s_\pi$ appears in $t^\one(\fba)$ instead of
  in $t^\nul(-\fba)$. Also,
$(-1)^{|\CD^{(0)}|}=1$, $(-1)^{|\CD^{(1)}|}=-(-1)^{\sig_0+\sig_\pi}$ in the case of even $n$.
In the following products, for $l$ odd there are $n/2$ values and
 for $l$ even we have $n/2-1$ values:
\bea \fl R^{(n\,{\rm even})}(\bdr')\;&=&
\left({\textstyle \prod_{l\in \CD}}((-1)^{\rho_l}\,r_l\,+\fba)\;
  -\;(-1)^{\sig_0+\sig_\pi}{\textstyle \prod_{l\in \CD}}((-1)^{\rho_l}\,r_l\,-\fba)\times\right.\ny\\ \fl
&&\hspace*{-1cm}\left.\times (a^4-1)\lk \fba\,+(-1)^{\sig_0}\lms_0\rk \;\lk \fba\,\,+
(-1)^{\sig_\pi}\lms_\pi\rk\  a^{2}/ (\aab \abb)\right)\:\RRR \,,
\label{rRe}\eea
\be \RRR\;=\;r_0^2\;  \aab^{-1} (\aab \abb)^{1-n/2} a^{n-2}\, \prod_{k\in \CDC}\; ((-1)^k\,\fba\,+r_k)\,.
\label{evenR}\ee

\subsection{Calculation of $\;\:\CN(\bdr)\cdot R_0(\bdr')$}

In this subsection we shall show that the product $\CN(\bdr)\cdot R_0(\bdr')$ can be put
into the very simple form \r{NRN}. Let us start evaluating $R_0(\bdr')$.

At the beginning of this Section we already discussed that, if we exclude case (iii), $l\neq\WD$, then
$\qs^{(0)}_l(0)\,\qs^{(1)}_l(0)\,=\,1$, and from (ii) if $l\in\CDC$
    we have $\qs^{(0)}_l(1)\,\qs^{(1)}_l(1)\,=\,0$.
So we have to consider only $l\in\CD$ for which we get from (i) and (iv)
(momentarily we suppose that $l$ is odd, but the result \r{QS} is the same for even $l$):
\bea\label{QQ}\fl
\qs^{(0)}_l(1)\,\qs^{(1)}_l(1)&=&\;-(-1)^{\sigma_{q_l}+l}
\frac{(-1)^{n}\,2\: {\rm i}\,\sin q_l\:{t}^{(0)}_{\check q_l}(-r_{l})}{n\,\chi_l\,r_{l}^{n-1}\:A(q_l)}\cdot
\frac{(-1)^{n-1}\:t^{(1)}(-r_{l})}{2\:\chi_l\: r_{l}^{n-1}\:F(r_{l})}\,\ny\\ &=&
(-1)^{\sigma_{q_l}+n+1}\frac{{\rm i}\,\sin{q_l}}{n\:A(q_l)\;F^{n-1}(r_l)}\;
     {t}^{(0)}_{\check q_l}(-r_{l})\;t^{(1)}(-r_{l})\,,
\eea
where in the last step we used \r{shchi}.
The polynomial ${t}^{(0)}_{\check q_l}(\lm)$ is $\:t^{(0)}(\lm)\:$ given by \r{tNS} with
  the factor $\:\lm^2-r_l^2$ omitted, see \r{tcheck}. In the first line the factor
  $-(-1)^{\sig_{q_l}+l}$ takes care whether
  $q_l$ or $-q_l$ is excited, the minus sign comes
  because $Q_l^\nul$ is a left eigenvector component.
Now since we should not use $l\in\CDC$ (if present, such a term leads to a
  vanishing contribution in \r{mat-res}), we have
\bea \fl{t}^{(0)}_{\check q_l}(-r_{l})\;t^{(1)}(-r_{l})&=&\frac{a^{4n}-1}{a^4-1}\;\SS_l
   \!\!\!\prod_{m\in\CD,m\neq l}(r_l^2-r_m^2)\ny\\ &=&\frac{a^{4n}-1}{a^4-1}\; \SS_l\!
   \prod_{k=1, \,k\neq l}^{n-1}(r_l^2\,-\,r_k^2)\;\prod_{k\in\CDC}\lk-\frac{-r_l+r_k}{r_l+r_k}\rk.\label{tthat}\eea
with $\SS_l:=\:(a^4-1)(-r_l+(-1)^{\sig_\pi}s_\pi)(-r_l+(-1)^{\sig_0}s_0)$.
Inserting \r{tthat} into \r{QQ} and using, recall \r{tcheck},
\be \fl r_{l}^2-r_{k}^2= 2\,(\cos{q_k} -\cos{q_l})\,F(r_l)/A(q_k)\,,\quad
{\textstyle \prod_{k=1}^{n-1}}\:A(q_k)=\frac{a^{4n}-1}{a^4\,-1}\:=\:\frac{r_0^2}{a^2-b^2}\,.  \label{rAk}\ee
and the trigonometric identity
\[\textstyle
2^{n-1}\:\sin^2 q_l\:\prod_{k\ne l} (\cos q_l - \cos q_k)=n (-1)^{l+1}
\]
we obtain
\be\label{QS}  \qs^{(0)}_l(1)\;\qs^{(1)}_l(1)\:=\:(-1)^{\sig_{q_l}+l+|\CDC|+1}\:\frac{\SS_l }
{ 2 {\rm i}\:\sin q_l\: F(r_l)}\:\prod_{k \in \CDC} \lk\frac{-r_l+r_k}{r_l+r_k} \rk,
\ee valid both for $q_l$ in R and for $q_l$ in NS.

Let us rewrite the ratio
$\SS_l\,/{ (2 {\rm i}\,\sin q_l\, F(r_{l}))}$
in a convenient way. By straightforward use of the definitions
\r{s},~\r{shchi} and \r{rzero} we find
\bea \fl\mp\frac{\SS_l}{2{\rm i}\:\sin q_l\: F(r_{l})}&=&
    \frac{-r_{l}+(-1)^{\sig_0}\alpha_{\pm q_l}}{r_{l}+(-1)^{\sig_0}\alpha_{\pm q_l}}\,,\quad
    \alpha_{q}=\frac{b^2 - e^{{\rm i}q}}{a^2 -e^{{\rm i}q}}\hq\mbox{for}\hx
    (-1)^{\sig_0}\,=\,+(-1)^{\sig_\pi}\,,\ny\\ \fl
\mp\frac{\SS_l}{2{\rm i}\:\sin q_l\: F(r_{l})}&=&
    \frac{-r_{l}+(-1)^{\sigma_0}\beta_{\pm q_l}} {r_{l}+(-1)^{\sigma_0}\beta_{\pm q_l}}\,,
    \quad\beta_{q}=\frac{b^2 e^{{\rm i}q}-1}{a^2 -e^{{\rm i}q}}\hq\mbox{for}\hx
    (-1)^{\sig_0}\,=\,-(-1)^{\sig_\pi}\,,\label{SSF}\eea
leading to (written such that it is valid for both $\rho_l=0\:$ and $\:\rho_l=1$):\\[-2mm]
\be  Q^\nul_l(\rho_l)\:Q^\one_l(\rho_l)\;=\; (-1)^{(n-1)\rho_l}\,\frac{(-1)^{\rho_l}r_{l}+ \xi_l}
{r_{l}\,+\,\xi_l}\cdot\prod_{k \in \CDC}\frac{(-1)^{\rho_l}r_{l}+r_{k}}{r_{l}+r_{k}}
  \,,\label{Rnu} \ee
 where
\be\fl \xi_l=\lb\,{\tilde\alpha_l\:=\:(-1)^{\sig_0}\:\alpha_{\tilde q_l}\atop
    \tilde\beta_l\:=\:(-1)^{\sigma_0}\:\beta_{\tilde q_l}}\quad  \mbox{for}
  \hq (-1)^{\sig_0}\,=\,\pm(-1)^{\sig_\pi}\,;\right.\quad
\tilde q_l=(-1)^{\sigma_{q_l}+|{\cal D}|+l}\,q_l\,.\label{xi}\ee

Multiplying by $\CN(\bdr')$, it is easy to see that the products over $k\in\CDC$ in \r{Rnu} cancel
(recall that $\rho_k=0$ for $k\in\CDC$)
and we get finally
\be \fl\CN(\bdr)\cdot R_0(\bdr')\,=\,\;\prod_{l\in\CD}(-1)^{\rho_l}\,\frac{(-1)^{\rho_l}r_{l}\,+\,\xi_l}
{r_{l}\,+\,\xi_l} \prod_{m\in\CD, m>l}\;
\frac{r_{l}+r_{m}}{(-1)^{\rho_{l}} r_{l}+ (-1)^{\rho_{m}}r_{m}}\,.\label{NRN}\ee

\subsection{Summation over $\bdr'$ in \r{mat-res}}\label{summ}

In \r{rRo}, \r{rRe} and \r{NRN} we have obtained all factors for the calculation of the normalized
matrix element in such a form that the dependence on the summation indices $\bdr'$ is explicit:
\bea\fl
\frac{\langle\, \Phi_0\,|\:\sz{n}\:|\,\Phi_1\,\rangle}
{\langle \,\tilde\Psi_{{\bf 0}}\,|\,\tilde\Psi_{{\bf 0}}\,\rangle}\;
 &=&\sum_{\bdr'\,\in\,{\mathbb Z}_2^{n-1}} \lk {\cal R}_+^\nu\;
   \prod_{l\in \CD}((-1)^{\rho_l}\,r_l\,+\fba) + {\cal R}_-^\nu\;
   \prod_{l\in \CD}((-1)^{\rho_l}\,r_l\,-\fba) \rk\;\times\ny\\
 \fl &&\times\;\prod_{l\in\CD}(-1)^{\rho_l}\,\frac{(-1)^{\rho_l}r_{l}\,+\,\xi_l}
{r_{l}\,+\,\xi_l} \prod_{m\in\CD, m>l}\;
\frac{r_{l}+r_{m}}{(-1)^{\rho_{l}} r_{l}+ (-1)^{\rho_{m}}r_{m}}\,,
  \label{matc1} \eea
where the $\bdr'$-independent factors ${\cal R}_\pm^\nu$ can be read off from \r{rRo}, \r{oddR} and \r{rRe}, \r{evenR}.
The superscript $\nu$ stands for $n$ odd and even, respectively.

Collecting all factors which depend on $\rho_l$, $l\in\mathcal{D}$,
the problem of performing the multiple summation
over $\rho_l$, $l\in\CD$, reduces to the calculating following sum (proved in the Appendix):
\be \fl Y_{\CD}\,
 =\sum_{\rho_{l},\,\,l\in\CD}\; \frac{\prod_{l\in\CD}\; (-1)^{\rho_{l}}\;
     (\,(-1)^{\rho_{l}} r_{l}\,+\, \xi_l)
\;\:(\,(-1)^{\rho_{l}} r_{l}\, +\,\fba)} {\prod_{l<m,\: l,m\in\CD}\;
     (\,(-1)^{\rho_{l}} r_{l}\:+\, (-1)^{\rho_{m}}r_{m})}\label{SUMnew}
\ee\[
\fl = \lb { c_\alpha\,(b\pm a)\lk \prod_{j\in\CD}\qq{j}\mp a b\rk\:
    {\displaystyle \frac{\prod_{l\in\CD}\: (2\,r_l/a)
  \:f_l^{(D-1)/2}\;g_l^{(D-3)/2}}{\prod_{l,m\in\CD,\,l<m}\;(\pm h_{l,m})}}\,,\quad
\xi_l=\pm\alpha_{\tilde q_l}\,,\quad D\;{\rm odd}}
  \atop
c_\beta\,(1\mp a b)\lk\pm a b \pj+1\rk
   {\displaystyle \frac{\prod_{l\in\CD}\:(2\,r_l/a)\,(f_l\,g_l)^{D/2-1}}
    {\prod_{l,m\in\CD,\,l<m}\:(\pm h_{l,m})}}\,,\quad
    \xi_l=\pm\beta_{\tilde q_l}\,,\quad D\;{\rm even}
  \right.\]
with $D=|{\cal D}|$,
\be\fl\qquad c_\alpha\:=\:\aab^{-(D-1)(D-3)/4}\:(-\abb)^{-(D-1)^2/4}\:\,,\hx
    c_\beta\:=\: (-\aab)^{-(D-2)D/4} \abb^{-(D-2)^2/4} \,,\ee
and we abbreviate
\bea \label{fgh}
 \lefteqn{f_l\:=\:a^2\,\qq{l}-1\,,\hs g_l\:=\:\qq{l}-a^2\,,\hs
   h_{l,m}\:=\:\qqq{l}{m}-1,}\ny\\[2mm]  && \aab\:=\:a^2-b^2,\hs\hx\;\; \abb\:=\:1-a^2b^2.\eea

In the calculation of the matrix element \r{mat-res},
we restrict ourselves to the case $\sig_0=\sig_\pi$ corresponding to $D$ odd. The case $\sig_0\ne \sig_\pi$
corresponding to $D$ even can be done similarly.
Also the two cases of even and odd parity of $n$ have to be considered separately.
For odd $n$,
taking into account \r{rRo}, \r{SUMnew} and using \r{s} with
\bea \fl \lefteqn{(-1)^{\sig_0}a (a^2+1) \lk -\fba\,\,+(-1)^{\sig_0}\lms_\pi\rk ((-1)^{\sig_0}a+b)
\lk{\textstyle \prod_{l\in {\cal D}}} e^{{\rm i} \tilde q_l}-(-1)^{\sig_0}ab \rk\;-}\ny\\[1.5mm] \fl
\qquad -(-1)^{\sig_0}a (a^2-1) \lk \fba\,+(-1)^{\sig_0}\lms_0\rk ((-1)^{\sig_0}a-b)
\lk{\textstyle \prod_{l\in {\cal D}}} e^{{\rm i} \tilde q_l}+(-1)^{\sig_0}ab \rk\;=\\[1.5mm]
=  2\,(-1)^{\sig_0} \aab\,(1-(-1)^{\sig_0}ab)\, \textstyle \prod_{l\in {\cal D}} e^{{\rm i} \tilde q_l}\,,
\ny\eea
we have finally:
\bea \fl\frac{\langle\, \Phi_0\,|\,\sz{n}\,|\,\Phi_1\,\rangle}
{\langle\: \tilde\Psi_{0,{\bf 0}}\,|\,\tilde\Psi_{0,{\bf 0}}\:\rangle}\;
         &=&\;r_0\,{\tilde c}_n\:c_\alpha\:\prod_{k\in\CDC}((-1)^k b +a \,r_k)\;\times\ny\\
   \fl && \hspace*{4mm}\times\; \frac{\prod_{l\in\CD}\,2\,r_l\,\qq{l}\:f_l^{(D-1)/2}g_l^{(D-3)/2}}
    {\prod_{l<m, {l,m\in\CD}} (-1)^{\sig_0} h_{l,m}}
 \:\frac{\prod_{l<m,\:l,m \in\CD}\;(r_{l}+r_{m})}
 {\prod_{l\in\CD}\:(r_l+(-1)^{\sig_0}\alpha_{\tilde q_l})}\,,\label{Rself}\eea
where
\[
{\tilde c}_n=((-1)^{\sig_0}-a\,b)\;(\aab\,\abb)^{(1-n)/2}\,.
\]

Analogously, in the case of even $n$ and $\sig_0=\sig_\pi$, using \r{rRo} we get the
same formula \r{Rself} for the matrix elements but with
\[{\tilde c}_n=(b+(-1)^{\sig_0}a)\; \aab^{-1} (\aab\,\abb)^{(2-n)/2}\,.\]

\section{Product of matrix elements}

In this section we sketch the calculation of the conjugate matrix elements
$\langle \Phi_1|\sz{n}|\Phi_0\rangle$, where the vectors $\langle \Phi_1|$  and $|\Phi_0\rangle$
shall have the same eigenvalues as the vectors $|\Phi_1\rangle$ and $\langle \Phi_0|$  used
in the previous sections. This calculation can be performed in the same way as we did in Section 4.
In analogy to \r{mat-res}, \r{Rs} we get for the homogeneous case
\[ \frac{\langle\, \Phi_0\,|\,\!\sz{n}\,\!|\,\Phi_1\,\rangle}
{\langle \,\tilde\Psi_{0,\bnul}\,|\,\tilde\Psi_{0,\bnul}\,\rangle}
\;=\;\frac{a}{2\,r_0}\;\sum_{\bdr'} \CN(\bdr')\; R_0^*(\bdr')\;R^*(\bdr')\,,
\]
with
\[
R^*(\bdr')\;=\;\frac{t^{(1)}(-\fba)}{\prod_{l=1}^{n-1} (-\fba+(-1)^{\rho_{l}} r_{l})}\;+\;
\frac{t^{(0)}(\fba)}{\prod_{l=1}^{n-1} (\fba+(-1)^{\rho_{l}} r_{l})}\,.
\]
Now we have to make the same transformations as we made for $R(\bdr')$ in Section \ref{calcR}.
The expression for $R^*(\bdr')$ is obtained from $R(\bdr')$
  given by \r{Rs} just by substituting $b\to -b$ (in particular, $\fba\to -\fba$).
Let us compare $R_0^*(\bdr')$ and $R_0(\bdr')$. From the solution of the Baxter equations it follows that
\[\fl\qquad Q^{\rL(1)}_l(\rho_{l})\;Q^{\rR(0)}_l(\rho_{l})\;=\;Q^{\rL(0)}_l(\rho_{l})\; Q^{\rR(1)}_l(\rho_{l})
   \hs\mbox{unless}\hx l\in \CD \hx\mbox{and}\hx \rho_{l}\,=1. \]
\[\fl\qquad Q^{\rL(1)}_l(1)\;Q^{\rR(0)}_l(1)\;=\;-\:Q^{\rL(0)}_l(1)\; Q^{\rR(1)}_l(1)
   \hs\mbox{if}\hx l\in \CD. \]
So, in the final formula we have to substitute $\;\tilde q_l\to -\tilde q_l$.

Using these rules, in the case of $\sigma_0=\sigma_\pi$, from \r{Rself}
we get
\bea \fl\lefteqn{\frac{\langle\, \Phi_1\,|\,\sz{n}\,|\,\Phi_0\,\rangle}
{\langle\: \tilde\Psi_{0,{\bf 0}}\,|\,\tilde\Psi_{0,{\bf 0}}\:\rangle}\;
         \;=\;r_0{\tilde c}_n c_\alpha\:
 \prod_{k\in\CDC}(-(-1)^k b\,+\,a\, r_k)\times}
         \ny\\ \fl &&\hspace*{2cm}
    \times \;\frac{\prod_{l\in\CD}\,2\,r_l\,e^{-{\rm i} \tilde q_l}\:(f_l^*)^{(D-1)/2}(g_l^*)^{(D-3)/2}}
    {\prod_{l<m,\, {l,m\in\CD}} (-1)^{\sig_0}h^*_{l,m}}\;
 \:\frac{\prod_{l<m,\:l,m \in\CD}\;(r_{l}+r_{m})}
 {\prod_{l\in\CD}\:(r_l+(-1)^{\sig_0}\alpha_{-\tilde q_l})},\label{Rconj}\eea
where $h^*_{l,m},\;f^*_l,\;g^*_l$ are $h_{l,m},\;f_l,\;g_l$ from \r{fgh}
with the replacement $\tilde q_l\to -\tilde q_l$.
    $\;\,\RRR^*\:$ is $\,\RRR\,$ with $\,b\to -b$.

In the product of \r{Rself} with \r{Rconj} nice simplifications appear.
In $\;\RRR\,\cdot\,\RRR^*$ we can use
\be \prod_{k\in \CDC}((-1)^k\,b+a r_k)\,(-(-1)^k\,b+a\,r_k)\,=\,\prod_{k\in \CDC}\frac{\aab\,\abb}{A({\tilde q}_l)}
  \,,\ee
and $f_l\cdot f_l^*\:=\:g_l\cdot g_l^*\:=\:A(\qt_l)$,
so that
\bea \fl \lefteqn{ \frac{\langle\, \Phi_0\,|\,\sz{n}\,|\,\Phi_1\,\rangle}
{\langle\: \tilde\Psi_{0,{\bf 0}}\,|\,\tilde\Psi_{0,{\bf 0}}\:\rangle}
      \frac{\langle\, \Phi_1\,|\,\sz{n}\,|\,\Phi_0\,\rangle}
{\langle\: \tilde\Psi_{0,{\bf 0}}\,|\,\tilde\Psi_{0,{\bf 0}}\:\rangle}
\;\:={\tilde c}_n {\tilde c}^*_n\;r_0^2\;(c_\alpha)^2\;\prod_{k\in\CDC}\;\frac{\aab\,\abb}{A(\qt_k)}\;\times}
    \ny \\ \fl \hspace*{2.5cm}\times\;\prod_{l\in\CD}\,\lk
    \frac{ 4\,r_l^2\:A(\qt_l)^{D-2}}{(r_l\,+\,(-1)^{\sig_0}\alpha_{\qt_l})\,(r_l\,+\,(-1)^{\sig_0}\alpha_{-\qt_l})}
    \:\prod_{m>l \atop {m\in\CD}}\frac{(r_l\,+\,r_m)^2}{|h_{l,m}|^2}\rk\!,
\label{PRO}\eea
where ${\tilde c}^*_n$ is ${\tilde c}_n$ with $b\to -b$ and ${\tilde c}_n {\tilde c}^*_n= \aab^{1-n} \,\abb^{2-n}$
is independent of parity of $n$.
Using further \r{rAk}
and the short notations: $\lm_0=(-1)^{\sigma_0}s_0$,
$\lm_\pi=(-1)^{\sigma_\pi}s_\pi$  with $\sig_0=\sig_\pi$:
\[  \fl
|h_{l,m}|^2=2\,(\cos \qt_m\,-\,\cos \qt_l)\:\:
\frac{\sin\frac{1}{2}(\tilde q_l+\tilde q_m)}{\sin\frac{1}{2}(\qt_l\,-\qt_m)}
   =\frac{r_l^2\,-\,r_m^2}{-\aab\:\abb}\;A(\qt_m)\;A(\qt_l)\;\;
   \frac{\sin\frac{1}{2}(\tilde q_l+\tilde q_m)}{\sin\frac{1}{2}(\qt_l\,-\qt_m)}\,,\]
\[ \fl \frac{\lm_\pi-\lm_0}{\lm_\pi+\lm_0}=-\frac{\aab}{\abb}\,,\quad
2r_l(r_{l}+\lm_0)(r_l+\lm_\pi)=
(\lm_0+\lm_\pi)(r_l+(-1)^{\sig_0}\alpha_{\qt_l})(r_l+(-1)^{\sig_0}\alpha_{-\qt_l})\,,
\]
we find that in \r{PRO} the factors $A(\qt_{l})$ combine to
\be \fl\prod_{k\in\CDC}\frac{1}{A(\qt_{k})}\;\;\prod_{l\in\CD}\;\lk\,A(\qt_l)^{D-2}
         \!\!\!\!\prod_{{m>l},\; {m\in\CD}}\frac{1}{A(\qt_m)\;A(\qt_l)}\rk\;=
         \;\prod_{l\in\CDC\cup\CD}\frac{1}{A(\qt_{l})}\;=\;\frac{\aab}{r_0^2}\,.\label{cancA}\ee
Since also the factors $\aab$ and $\abb$ in \r{PRO} can be collected as follows ($|\CDC|+D=n-1$):
\[ {\tilde c}_n {\tilde c}^*_n (c_\alpha)^2
         (\aab\,\abb)^{|\CDC|}(-\aab\,\abb)^{D(D-1)/2}\aab\;=\;(-\aab/\abb)^{(D-1)/2}\,,\]
we get finally for arbitrary $n$ and ${\sigma_0}={\sigma_\pi}$
\bea\fl
\lefteqn{
\frac{\langle\, \Phi_0\,|\,\sz{n}\,|\,\Phi_1\,\rangle\;\langle\, \Phi_1\,|\,\sz{n}\,|\,\Phi_0\,\rangle}
{\langle \,\tilde\Psi_{0,{\bf 0}}\,|\,\tilde\Psi_{0,{\bf 0}}\,\rangle^2}
\:=\:(\lm_\pi^2\,-\,\lm_0^2\,)^{(D-\delta)/2} \:(\lm_0\,+\lm_\pi)^{\delta}\:
\prod_{l \in \CD} \frac{2\,r_l}{(\lm_0\,+r_l)\,(\lm_\pi\,+r_l)}\;\times}\ny\\
 &&\times
\prod_{l<m,\,\, l,m \in\CD}\frac{r_l\,+r_m}{r_l\,-r_{m}}\:\cdot\:
\frac{\sin\,\frac{1}{2}(\qt_l-\qt_m)}{\sin\,\frac{1}{2}(\qt_l\,+\qt_m)}
\label{ME}\eea
where $\delta=1$. In a similar way we can find the product of matrix elements in the case of
${\sigma_0}\ne {\sigma_\pi}$. The final result is \r{ME} with $\delta=0$.
Observe that using \r{ME}, the explicit appearance of excitations of
type (ii), i.e. $k\in\CDC$ has disappeared from our formula (recall that we still exclude $k\in\WD$).
We still have normalized our matrix elements by the norm taken from the auxiliary system. In the next section we shall
normalize to the norm of periodic states so that the spin matrix element becomes independent of the special
normalization of the periodic states chosen. We shall also include the hitherto excluded case (iii).

\section{Final formula for the square of the matrix element}\label{Final}

The proper quantity to consider for the matrix element of the spin operator,
which does not depend on the normalization of the
eigenvectors of the transfer-matrix, is
\be\label{px}\langle \Phi_0|\sz{n}|\Phi_1\rangle\langle \Phi_1|\sz{n}|\Phi_0\rangle/
(\langle \Phi_0|\Phi_0\rangle\langle \Phi_1|\Phi_1\rangle)\,.\ee
The factor
\be
\langle \Phi_0|\Phi_0 \rangle \langle \Phi_1|\Phi_1\rangle/
\langle \tilde\Psi_{0,\bnul}|\tilde\Psi_{0,\bnul}\rangle^2
\label{chn}\ee
providing this change of normalization has been derived
in formulas (74) and (75) of \cite{gipst1} for odd and even $n$, respectively.
To obtain \r{px} we just divide \r{ME} by \r{chn}.
The final result for the square of the matrix element \r{ME} as well as the formulas for the
squares of norm \r{chn} were obtained for the eigenvectors with eigenvalues
not containing factors $(\lm-\lms_{q_k})^2$, i.e. up to now we have excluded the case (iii)
in Section \ref{zweizwei}.

\subsection{Excitations $j\in\WD$ producing factors $(\lm-\lms_{q_j})^2$ in $\;t^\rho(\lm)$.}

Now we explain how to modify these formulas for the matrix elements and norms
if the eigenvalue polynomials of states $|\Phi_0\rangle$ and $|\Phi_1\rangle$ contain factors
$(\lm-\lms_{q_j})^2$ for some $j$, so that $t^\rho(r_j)=0$ for $\rho$ satisfying
$(-1)^\rho=(-1)^{j-1}$.
We denote the set of such $j$, corresponding to both states $|\Phi_0\rangle$ and $|\Phi_1\rangle$,
by $\WD$.
As already mentioned in Section~\ref{zweizwei}, case (iii),
we cannot normalize the solution to the Baxter equation for the state $|\Phi_\rho\rangle$ by
$Q^{\rm L,R(\rho)}_j(0)=1$. However, we may normalize it by $Q^{\rm L,R(\rho)}_j(1)=1$.
For the other state, the solution to the Baxter equation is managed by the case (i) and there is no problem
with the normalization $Q^{\rm L,R(\rho)}_j(0)=1$. We shall use the normalization
$Q^{\rm L,R(\rho)}_j(1)=1$ for all $j\in \WD$ irrespective of whether we have case (i) or (iii).

In \cite{gipst1} the norm (56) and the matrix element \r{mat-res} was calculated using the  normalization
$Q^{\rm L,R(\rho)}_j(0)=1$ for all $j=1,2,\ldots,n-1$. These formulas were obtained
for the case of generic parameters for which any normalization is possible.
Let us trace the changes in these formulas if instead we choose the normalization
$Q^{\rm L,R(\rho)}_j(1)=1$ for $j$ from a subset $\WD\subset \{1,2,\ldots,n-1\}$.

Compare the formulas for the matrix element \r{mat-res} corresponding to
a different normalization of the solutions to the Baxter equation,
i.e. we compare the term corresponding to a set $\bdr'$ in one formula with the term
corresponding to the set ${\bdr'}^{+\WD}$ in the other formula
(the set ${\bdr'}^{+\WD}$ is obtained from the set $\bdr'$
by shifts $\rho_j\to\rho_j+1$, $j\in\WD$).
For $N=2$ this just means interchanging $\rho_j=0$ and $\rho_j=1$,
while the other components remain unchanged.
Also, we change all $r_j\to -r_j$, $j\in\WD$ in the second formula.

{}From the Baxter equations \r{Bax}, \r{BaxL},
the solution $Q^{\rm L,R(\rho)}_j(\rho_j+1)$ normalized by $Q^{\rm L,R(\rho)}_j(1)=1$ coincides with
$\left.Q^{\rm L,R(\rho)}_j(\rho_j)\right|_{r_j\to-r_j}$ with the normalization $Q^{\rm L,R(\rho)}_j(0)=1$.
The only factor which is changing
in \r{mat-res} is $\CN(\bdr')$. The denominator is unchanged under the simultaneous substitutions
$r_j\to-r_j$ and $\rho_j\to\rho_j+1$. The change in the numerator is corrected by
the division of the matrix elements corresponding to solutions of Baxter equations with $Q^{\rm L,R(\rho)}_k(1)=1$,
not by $\langle \tilde\Psi_{0,{\bf 0}}|\tilde\Psi_{0,{\bf 0}}\rangle$ but by
$\langle\tilde\Psi_{0,\bnul^{+\WD}}|\tilde\Psi_{0,\bnul^{+\WD}}\rangle$.
{}From the general expression for the norm, (20) of \cite{gipst1} at $N=2$,
the change of normalization means multiplying our matrix elements by
\be  \frac{\langle \,\tilde\Psi_{0,\bnul}\,|\,\tilde\Psi_{0,\bnul}\,\rangle}
     {\langle\tilde\Psi_{0,\bnul^{+\WD}}|\tilde\Psi_{0,\bnul^{+\WD}}\rangle}\;=\;
\frac{\prod_{l<m}^{n-1} (r_m+r_l\,)}
{\prod_{l<m}^{n-1} (r_m (-1)^{\rho_m}+r_l(-1)^{\rho_l})}.\ee
where $\rho_m=1$ if $m\in\WD$ and $\rho_m=0$ otherwise.
Finally, the factor $(-1)^{n\tilde\rho'}$ gives the sign $(-1)^{|\WD|n}$.
The formula for the norms undergoes the same changes.
Therefore formally the expression \r{px} has to be invariant with respect to the substitutions $r_j\to-r_j$.
But the final formula for \r{px} is given for the case of Ising model
where we already replaced the dependence on $s_{q_j}$ using the coincidence with $r_j$.
So the substitutions $r_j\to-r_j$ in this final expression for \r{px}
given in terms of $\lm_0$, $\lm_\pi$ and $r_{k}$, $k=1,2,\ldots,n-1$,
will also change the eigenvectors entering \r{px}
to the eigenvectors corresponding to eigenvalue polynomials with factors $(\lm-r_j)^2$, $\,j\in \WD$,
instead of $(\lm+r_j)^2$. This is exactly what we need.

Summarizing:
The final result \r{px} was obtained for eigenvectors with eigenvalues not containing
factors $(\lm-\lms_{q_j})^2$ and which could be given in terms of $\lm_0$, $\lm_\pi$ and $r_{k}$, $k=1,2,\ldots,n-1$.
Now, if the eigenvalue polynomial contains the factors
$(\lm-\lms_{q_j})^2$ instead of $(\lm+\lms_{q_j})^2$ for some $j$,
we just have to replace $r_{j}\to -r_{j}$ in the final formula for all such $j$.

\subsection{Final result in terms of $\;\:\lm_0$, $\lm_\pi$, $r_{k}$, and $\,\qt_l$.}

So the final formula for the matrix element becomes\\[-4mm]
\[\fl
\frac{\langle \Phi_0|\sz{n}|\Phi_1\rangle
\langle \Phi_1|\sz{n}|\Phi_0\rangle}
{\langle \Phi_0|\Phi_0\rangle\langle \Phi_1|\Phi_1\rangle}
\:=(\lm_\pi^2-\lm_0^2)^{(D-\delta)/2} (\lm_0+\lm_\pi)^{\delta}
\prod_{{l<m}\atop{l,m \in \CD}}\!\!\lk
\frac{r_{l}+r_{m}}{r_{l}-r_{m}}\cdot\frac{\sin\,\frac{1}{2}(\qt_l-\qt_m)}{\sin\,\frac{1}{2}(\qt_l\,+\qt_m)}\rk
\times
\]
\be\fl\times\frac{\Lambda_n}{2^D \prod_{k\in\overline {\cal D}} (\dot + 2r_{k})}\cdot\frac
{\prod_{k{\rm \, odd},\, l{\rm \,even}}  \Bigl((\dot - r_{k}\dot +r_{l}) (\dot+r_{k}\dot- r_{l})\Bigr)}
{ \prod_{k<l, k,l{\rm \, odd}} \Bigl((\dot + r_{k} \dot +r_{l}) (\dot - r_{k}\dot -r_{l})\Bigr)
\prod_{k<l, k,l{\rm \, even}} \Bigl((\dot + r_{k}\dot+r_{l}) (\dot- r_{k}\dot- r_{l})\Bigr)},\label{prq}
\ee
where
\[\fl
\Lambda_n=\frac{\prod_{k\in \overline {\cal D}^\nul} (\lm_0\dot +r_k)}
{\prod_{k\in \overline {\cal D}^\one} (\lm_0\dot +r_k)\prod_{k\in\CD^\one} (\lm_0^2-r_k^2)}
\cdot
\frac{\prod_{k\in \overline {\cal D}^\one} (\lm_\pi\dot+r_k)}
{\prod_{k\in \overline {\cal D}^\nul} (\lm_\pi\dot+r_k)\prod_{k\in\CD^\nul} (\lm_\pi^2-r_k^2)}\,,\
\mbox{for odd $n$},
\]
\[\fl
\Lambda_n=\frac{\prod_{k\in \overline {\cal D}^\nul} (\lm_0\dot +r_k)(\lm_\pi\dot+r_k)}
{(\lm_0+\lm_\pi)\prod_{k\in \overline {\cal D}^\one} (\lm_0\dot +r_k)(\lm_\pi\dot+r_k)
\prod_{k\in\CD^\one} (\lm_0^2-r_k^2)(\lm_\pi^2-r_k^2)}\,,\qquad
\mbox{for even $n$},
\]
$\overline {\cal D}= \CDC\cup\WD$, $\overline {\cal D}^\nul=\CDC^\nul\cup \WD^\nul$,
$\overline {\cal D}^\one=\CDC^\one\cup \WD^\one$, and
$\dot \pm r_m$ is the short notation for $r_m$ if $m\in \CDC$, $\pm r_m$ if $m \in \CD$ and
$-r_m$ if $m\in \WD$, respectively.
The right hand side of the first line is just the right hand side of \r{ME} except for the
last product in the first line of \r{ME}, which is partly cancelled by the terms in the first
line of the change of normalization \r{chn} taking from (74) and (75) of \cite{gipst1}.
Also, we have taken into account the contributions of $k\in\WD$:
these have the same form as those for $k\in\CDC$, just with reflected $r_k\to -r_k$ as discussed above.

We claim that these formulas prove the formula for the matrix element which was given in \cite{BL2}
in an equivalent version.
In order to show the equivalence, in the following we perform some transformations of \r{prq}
to get a formula which is more appropriate for comparison.

\subsection{Final result in terms of momenta}

Let $\{\qu_1$, $\qu_2,\ldots,$ $\qu_K\}$ and $\{\pu_1$, $\pu_2,\ldots$, $\pu_L\}$
be the sets of the momenta of the excitations presenting the states $|\Phi_0\rangle$
from the NS-sector and $|\Phi_1\rangle$ from the R-sector, respectively.
After some lengthy but straightforward
transformations of \r{prq} we obtain
\[\fl
\frac{\langle\, \Phi_0\,|\,\sz{n}\,|\,\Phi_1\,\rangle  \langle\,\Phi_1\,|\,\sz{n}\,|\,\Phi_0\,\rangle}
{\langle\, \Phi_0\,|\,\Phi_0\,\rangle \,\langle\,\Phi_1\,|\,\Phi_1\,\rangle}
\;=\; J \:(\lms_\pi+\lms_0)\, (\lms_\pi^2-\lms_0^2)^{(K+L-1)/2}\;\times\]
\be\fl \qquad\times\;
\prod_{k=1}^K
\frac{P^{\rm NS}_{\qu_k}\prod_{\qu\ne |\qu_k|}^{\frac{{\rm NS}}{2}} N_{\qu,\qu_k}}
{\prod_{\pu}^{\frac{{\rm R}}{2}} N_{\pu,\qu_k}}\cdot
\prod_{l=1}^L
\frac{P^{\rm R}_{\pu_l}\prod_{\pu\ne |\pu_l|}^{\frac{{\rm R}}{2}} N_{\pu,\pu_l}}
{\prod_{\qu}^{\frac{{\rm NS}}{2}} N_{\qu,\pu_l}}\cdot
\frac{\prod_{k=1}^K\prod_{l=1}^L M_{\qu_k,\pu_l}}
{ \prod_{k<k'}^K M_{\qu_k,\qu_{k'}}
\prod_{l<l'}^L M_{\pu_l,\pu_{l'}}}\,, \label{MEs}
\ee
where NS/2 (R/2) is the subset of quasi-momenta from NS (R) taking values in the segment $0<\qu<\pi$\,,~
NS/2 (R/2) containing $q_k$ with odd $k$ (even $k$):
\[ M_{\alpha,\beta} = \frac{\lms_\alpha+\lms_\beta}{\lms_\alpha-\lms_\beta}
\cdot\frac{\sin\frac{\alpha+\beta}{2}}{\sin\frac{\alpha-\beta}{2}}\,,\qquad
M_{\alpha,-\alpha} = \frac{\lms_\alpha^2\:
(\lms_0^2-\lms_\pi^2)}{(\lms_\pi^2-\lms_\alpha^2)(\lms_0^2-\lms_\alpha^2)}\,,
\]
\[
N_{\alpha,\beta} = \frac{\lms_\alpha+\lms_\beta}{\lms_\alpha-\lms_\beta},\hq {\cal J}\,=\:
\frac{{\prod_{\qu}}^{\!\!\!\frac{{\rm NS}}{2}}\:(\lms_0+\lms_\qu)}
  {{\prod_{\pu}}^{\!\!\!\frac{{\rm R}}{2}}\: (\lms_0+\lms_\pu)}\cdot
\frac{\prod_{\qu}^{\frac{{\rm NS}}{2}}\prod_{\pu}^{\frac{{\rm R}}{2}} (\lms_\qu+ \lms_\pu)^2 }
{ \prod_{\qu,\qu'}^{\frac{{\rm NS}}{2}}( \lms_\qu+ \lms_{\qu'})
\prod_{\pu,\pu'}^{\frac{{\rm R}}{2}} (\lms_\pu+ \lms_{\pu'})}\, .\]
For $n$ odd:\\[-11mm]
\[
P_{\qu}^{\rm NS}= \frac{\lms_\qu}{(\lms_\pi-\lms_\qu)(\lms_0+\lms_\qu)},\hx\qu\neq \pi,\hx
P_{\pu}^{\rm R} = \frac{\lms_\pu}{(\lms_\pi+\lms_\pu)(\lms_0-\lms_\pu)},\hx\pu\neq 0,\]
\[P_{0}^{\rm R} = P_{\pi}^{\rm NS} = \frac{1}{\lms_\pi+\lms_0},\hs
 J =\frac{{\prod_{\pu}}^{\!\!\!\frac{{\rm R}}{2}}\;(\lms_\pi+\lms_\pu) }
 {{\prod_{\qu}}^{\!\!\!\frac{{\rm NS}}{2}}\;(\lms_\pi+\lms_\qu)}\;\:{\cal J}
 ,\]
for $n$ even:\\[-11mm]
\[
P_{\qu}^{\rm NS} = \frac{\lms_\qu}{(\lms_\pi+\lms_\qu)(\lms_0+\lms_\qu)},\quad
P_{\pu}^{\rm R} = \frac{\lms_\pu}{(\lms_\pi-\lms_\pu)(\lms_0-\lms_\pu)},\hx\pu\neq 0,\,\pi,\]
\[ P_{0}^{\rm R} = -P_{\pi}^{\rm NS} = \frac{1}{\lms_\pi-\lms_0},\hs
 J =\frac{{\prod_{\pu}}^{\!\!\!\frac{{\rm NS}}{2}}\;(\lms_\pi+\lms_\qu) }
 {{\prod_{\qu}}^{\!\!\!\frac{{\rm R}}{2}}\;\:(\lms_\pi+\lms_\pu)}\;\:{\cal J}\,.\]

\section{Bugrij--Lisovyy formula for matrix element}

In ~\cite{BL1,BL2}, the matrix elements of $\sigma^z_k$
between eigenvectors of {\it symmetric} Ising transfer-matrix
\be\label{IsingSym}\fl
{\bf t}_{\rm IsingSym}=
\exp{ \lk\half{\textstyle \sum_{k=1}^n}\, K_x \,\sigma^z_{k-1}\,\sigma^z_k\rk}\;\;
\exp{\lk {\textstyle \sum_{k=1}^n}\, K^*_x\, \sigma^x_k\rk}\;\;
\exp{\lk\half{\textstyle \sum_{k=1}^n}\, K_x \,\sigma^z_{k-1}\,\sigma^z_k\rk}\,.
\ee
were given. Since \r{Isitra} and \r{IsingSym} are related by a similarity transformation with
\[\exp{ \lk\half\;{\textstyle \sum_{k=1}^n}\, K_x \,\sigma^z_{k-1}\,\sigma^z_k\rk},\]
which commutes with  $\sigma^z_m$, it is natural
to compare \r{MEs} with the square of the matrix element as given in \cite{BL2}:
\[\fl
|\,{}_{\rm NS}\langle\, \qu_1, \qu_2,\ldots,\qu_K\,|\;\sigma^z_m\;|\,\pu_1,\pu_2,\ldots,\pu_L\,\rangle_{\rm R}|^2=
\]\[\fl
\hq=\;\xi\; \xi_T\; \prod_{k=1}^K \;\frac{\prod^{\rm NS}_{\qu\ne \qu_k} \sinh \frac{\g(\qu_k)+\g(q)}{2}}
{n \prod^{\rm R}_{\pu} \sinh \frac{\g(\qu_k)+\g(\pu)}{2}}\;\;
\prod_{l=1}^L \;\frac{\prod^{\rm R}_{\pu\ne \pu_l} \sinh \frac{\g(\pu_l)+\g(\pu)}{2}}
{n \prod^{\rm NS}_{\qu} \sinh \frac{\g(\pu_l)+\g(\qu)}{2}}\cdot
\left(\frac{t_y-t_y^{-1}}{t_x-t_x^{-1}}\right)^{\!\!(K-L)^2/2}\!\!\!\times
\]\be\fl\qquad\hq \times\;
\prod_{k<k'}^K \frac{\sin^2\frac{\qu_k-\qu_{k'}}{2}} {\sinh^2 \frac{\g(\qu_k)+\g(\qu_{k'})}{2}}
\;\;\prod_{l<l'}^L \frac{\sin^2\frac{\pu_l-\pu_{l'}}{2}} {\sinh^2 \frac{\g(\pu_l)+\g(\pu_{l'})}{2}}
\prod_{1\le k \le K \atop 1\le l \le L}
\frac {\sinh^2 \frac{\g(\qu_k)+\g(\pu_l)}{2}} {\sin^2\frac{\qu_k-\pu_l}{2}}\,.
\label{ME_BL}
\ee
In this formula the states are labelled by the momenta of the excitations,
and the squared matrix element is given for
$\sigma^z_m$, $m=1,\ldots,n$. The operators $\sigma^z_m$ with
different values of $m$ are related by similarity transformations with the translation operator.
The states under consideration are eigenvectors of the translation operator \cite{Lisovyy}
with eigenvalues which have unit absolute value. This explains why the formula presented does not depend on $m$.
Therefore it is sufficient to calculate the matrix element for
$\sigma^z_n$. The factors in front of the right hand side of \r{ME_BL} are defined by
\[\fl
\xi=((\sinh 2K_x \sinh 2K_y)^{-2}-1)^{1/4},\hx
\xi_T=\left(
\frac{\prod^{\rm NS}_{\qu} \prod^{\rm R}_{\pu} \sinh^2 \frac{\g(\qu)+\g(\pu)}{2}}
{\prod^{\rm NS}_{\qu,\qu'} \sinh \frac{\g(\qu)+\g(\qu')}{2}
\prod^{\rm R}_{\pu,\pu'} \sinh \frac{\g(\pu)+\g(\pu')}{2}} \right)^{1/4}\!\!,
\]
where $\gamma(\qu)$ is the energy of the excitation with quasi-momentum $\qu$:
\be\label{dr}\cosh \gamma(\qu)=\frac{(t_x+t_x^{-1})(t_y+t_y^{-1})}
{2(t_x^{-1}-t_x)}-\frac{t_y-t_y^{-1}}{t_x-t_x^{-1}}\,\cos \qu\,,
\ee
where by \r{KKIsing} $\;t_x\,=\,\tanh{K_x}\;=\,a\,b$, $\;\:t_y\,=\,\tanh{K_y}\,=\,(a-b)/(a+b)$.

The excitation with quasi-momentum $\qu$ leads to the multiplication of the transfer matrix eigenvalue
by $e^{-\gamma(\qu)}$ in the notations of \cite{BL2},
and in our notation to the multiplication by $\pm(\lm-\lms_\qu)/(\lm+\lms_\qu)$ at $\lm=b/a$.
 The sign is not fixed because the excitations arise by pairs.
Comparing \r{dr} and \r{tmgen} we get
\be\label{gs}
e^{\gamma(\qu)}=\frac{a\, \lms_\qu\,+b}{a\, \lms_\qu\,-b}\,.
\ee
Therefore
\be\label{gabs}
\sinh\half(\gamma(\alpha)+\gamma(\beta))\:=\:e^{\frac{1}{2}(\gamma(\alpha)+\gamma(\beta))}
\frac{a\,b\,(\lms_\alpha+\lms_\beta)}{(b+a\lms_\alpha)(b+a\lms_\beta)}\,,
\ee
which leads, in particular, to
\be\label{4a}
\frac{\sinh\frac{\gamma(\alpha_1)+\gamma(\alpha_2)}{2}\cdot
\sinh\frac{\gamma(\alpha_3)+\gamma(\alpha_4)}{2}}
{\sinh\frac{\gamma(\alpha_1)+\gamma(\alpha_3)}{2}\cdot
\sinh\frac{\gamma(\alpha_2)+\gamma(\alpha_4)}{2}}=
\frac{(\lms_{\alpha_1}+\lms_{\alpha_2})(\lms_{\alpha_3}+\lms_{\alpha_4})}
{(\lms_{\alpha_1}+\lms_{\alpha_3})(\lms_{\alpha_2}+\lms_{\alpha_4})}\,.
\ee
Our next problem is to rewrite \r{ME_BL} in terms of $\lms_\qu$. We need
\[
\xi \left(\frac{\sinh^2\frac{\gamma(0)+\gamma(\pi)}{2}}{\sinh \gamma(0)\sinh\gamma(\pi)}\right)^{1/4}=
\frac{1}{(\sinh 2K_x \sinh 2K_y)^{1/2}}\,,
\]\[
\frac{t_y-t_y^{-1}}{t_x-t_x^{-1}}=\frac{\sinh 2K_x}{\sinh 2K_y}\,,\qquad
\sinh\frac{\gamma(0)+\gamma(\pi)}{2}=\frac{1}{\sinh 2K_y}\,.
\]
The following formulas give a correspondence between different parts of \r{ME_BL} and \r{MEs}:
\[
\frac{\xi\: \xi_T}{\sih{0}{\pi}}
\left(\frac{t_y-t_y^{-1}}{t_x-t_x^{-1}}\right)^{1/2}=J\,,
\]\[
\frac{\prod^{\rm NS}_{\qu\ne \qu_k} \sinh \frac{\g(\qu_k)+\g(\qu)}{2}}
{n \prod^{\rm R}_{\pu} \sinh \frac{\g(\qu_k)+\g(\pu)}{2}}=
\frac{\lms_0+\lms_\pi}{\sinh \frac{\g(0)+\g(\pi)}{2}}
\frac{P^{\rm NS}_{\qu_k}\prod_{\qu\ne |\qu_k|}^{\frac{{\rm NS}}{2}} N_{\qu,\qu_k}}
{\prod_{\pu}^{\frac{{\rm R}}{2}} N_{\pu,\qu_k}}\,,
\]\[
\frac{\prod^{\rm R}_{\pu\ne \pu_l} \sinh \frac{\g(\pu_l)+\g(\pu)}{2}}
{n \prod^{\rm NS}_{\qu} \sinh \frac{\g(\pu_j)+\g(\qu)}{2}}=
\frac{\lms_0+\lms_\pi}{\sinh \frac{\g(0)+\g(\pi)}{2}}
\frac{P^{\rm R}_{\pu_l}\prod_{\pu\ne |\pu_l|}^{\frac{{\rm R}}{2}} N_{\pu,\pu_l}}
{\prod_{\qu}^{\frac{{\rm NS}}{2}} N_{\qu,\pu_l}}\,,
\]
where in the last two formulas we used
\be\label{sab}\fl
\sinh^2\half(\g(\alpha)+\g(\beta))\,=-\frac{t_y-t_y^{-1}}{t_x-t_x^{-1}}\cdot
\frac{\lms_\alpha+\lms_\beta}{\lms_\alpha-\lms_\beta}\,\cdot
\sin\half(\alpha-\beta)\cdot \sin\half(\alpha+\beta)\,,
\ee
and, in particular,
\be\label{g0pi}
\sinh^2\half(\g(0)+\g(\pi))\;=\;\frac{t_y-t_y^{-1}}{t_x-t_x^{-1}}\cdot
\frac{\lms_0+\lms_\pi}{\lms_0-\lms_\pi}\,,
\ee
together with \r{4a} and some trigonometric identities.
The formula \r{sab} also gives
\be\label{gM}
\frac{\sinh^2\frac{\g(\alpha)+\g(\beta)}{2}}{\sin^2\frac{\alpha-\beta}{2}}\;=\;
-\frac{t_y-t_y^{-1}}{t_x-t_x^{-1}}\;\: M_{\alpha,\beta}\,.
\ee
Formula \r{gM} is also valid for $\beta=-\alpha$, but in this case we have to use formulas
\r{gs} and \r{gabs}.
Finally, if we take into account \r{g0pi} and
that $K$ in \r{MEs} is even, $L$ is odd, then it is easy to see that the formulas \r{MEs} and \r{ME_BL} coincide.

\section{Matrix elements for the quantum Ising chain in a transverse field}

In this section we apply the formulas for the matrix elements obtained in Section~6 to the derivation of
the matrix elements for the quantum Ising chain in a transverse field.
Let us start from the $L$-operator \r{L-Ising} with $a=g^{-1/2}$ and $b=0$:
\be\label{L-IsChain}
L_k(\lm)=\lk\ba{cc}
1+\lm\,\sx{k} &  \lm \,g^{-1/2} \sz{k}\\
g^{-1/2} \sz{k}& \lm g^{-1} \ea \rk\,.
\ee
Expanding the transfer-matrix for the monodromy matrix \r{mm} with this $L$-operator we have:
\[{\bf t}_n(\lm) ={\bf 1}-\frac{2\lm}{g}\, \hat {\cal H}\,+\cdots\,, \qquad
 \hat {\cal H}=-\frac{1}{2}\sum_{k=1}^n (\sz{k} \sz{k+1}\,+\,g\, \sx{k})\,,
\]
where $\hat {\cal H}$
is the Hamiltonian of the periodic quantum Ising chain in a transverse field. {}From
\r{tmgen} we get the spectrum of this Hamiltonian:
\be\label{energy}
{\cal E}=-\frac{1}{2}\sum_\qu\pm\; {\ve(\qu)}
\ee
where the energies of the quasi-particle excitations are
\[\fl
\ve(\qu)=(1-2g\cos\qu+g^2)^{1/2}=\left((g-1)^2
+ 4g\sin^2\frac{\qu}{2 }\right)^{1/2},\qquad \qu\ne 0,\pi\,,\]
\[\ve(0)=g-1,\quad\quad \ve(\pi)=g+1\,.\]
In \r{energy}, the sign $+/-$ in the front of  ${\ve(\qu)}$ corresponds
to the absence/presence of the excitation with the momentum $\qu$.
The NS-sector includes the states with an even number of excitations, the R-sector includes
the states with an odd number of excitations.
The momentum $\qu$ runs over the same set as in \r{tmgen}.
Then the formula for matrix elements for $\sz{n}$ is given by \r{MEs}
with $\lms_\qu=g/\ve(\qu)$. After some simplification we get the analogue
of \r{ME_BL} now for the quantum Ising chain:
\[\fl
|{}_{\rm NS}\langle \qu_1,
\qu_2,\ldots,\qu_K|\sigma^z_m|\pu_1,\pu_2,\ldots,\pu_L\rangle_{\rm
R}|^2=  g^{\frac{(K-L)^2}{2}}\xi \xi_T \prod_{k=1}^K \frac{ \e^{\eta(\qu_k)}}{ n \ve(\qu_k)}\prod_{l=1}^L
\frac{ \e^{-\eta(\pu_l)}}{ n \ve(\pu_l)}
\times
\]
\be\fl \qquad\times
\prod_{k<k'}^{K}\left(\frac{2\sin \frac{\qu_k-\qu_{k'}}{2}}{\ve(\qu_k)+\ve(\qu_{k'})}\right)^2
\prod_{l<l'}^{L}\left(\frac{2\sin \frac{\pu_l-\pu_{l'}}{2}}{\ve(\pu_l)+\ve(\pu_{l'})}\right)^2
\prod_{k=1}^{K}\prod_{l=1}^{L}
\left(\frac{\ve(\pu_l)+\ve(\qu_k)}{2\sin \frac{\pu_l-\qu_k}{2}}\right)^2\,,\label{ME-IsCh}
\ee
where
\[
\xi= \left(g^2-1\right)^{\frac 1 4}\,,\quad
\xi_T=\frac{\prod_{\qu}^ {\rm NS} \prod_{\pu}^ {\rm R}
(\ve(\qu)+\ve(\pu))^{\frac 1 2}}{\prod_{\qu,\qu'}^{\rm NS}(
\ve(\qu)+ \ve(\qu'))^{\frac 1 4}\prod_{\pu,\pu'}^{\rm R} (\ve(\pu)+
\ve(\pu'))^{\frac 1 4}}
\]
and
\[
\e^{\eta(\qu)}=\frac{ \prod_{\qu' }^{ \rm NS}\left(
\ve(\qu)+\ve(\qu')\right)}{\prod_{\pu}^{ \rm R} \left(\ve(\qu
)+\ve(\pu)\right)}\,.
\]
Formally, all these formulas are correct for the paramagnetic phase where $g>1$
and for the ferromagnetic phase where $0\le g<1$. But for the case $0\le g<1$ it is natural
to redefine the energy of zero-momentum excitation as $\ve(0)=1-g$ to be positive.
{}From \r{energy}, this change of the sign of $\ve(0)$ in the ferromagnetic phase
leads to a formal change between absence-presence of zero-momentum excitation
in the labelling of eigenstates. Therefore
the number of the excitations in each sector (NS and R) becomes even.
Direct calculation shows that the change of the sign of $\ve(0)$ in \r{ME-IsCh}
can be absorbed to obtain formally the same formula \r{ME-IsCh} but with
new $\ve(0)$, even $L$ (the number of the excitations in R-sector) and
new $\xi=(1-g^2)^{1/4}$.

Formulas \r{ME_BL} and \r{ME-IsCh} allow to reobtain
already known formulas for the Ising model, e.g.
the spontaneous magnetization \cite{Onsager,Yang}.
Indeed, for the quantum Ising chain in the ferromagnetic phase
($0\le g<1$) and in the thermodynamic limit $n\to\infty$
(when the energies of $|\mbox{vac}\rangle_{\rm NS}$ and $|\mbox{vac}\rangle_{\rm R}$ coincide giving
the degeneration of the ground state), we have $\xi_T\to 1$
and therefore the spontaneous magnetization
${}_{\rm NS}\langle \mbox{vac}|\sigma^z_m|\mbox{vac}\rangle_{\rm R}=\xi^{1/2}=(1-g^2)^{1/8}$.


\section{Conclusions}

In this paper we calculated the normalized spin matrix element between arbitrary states of the Ising model,
the main result being the formulas \r{prq},\r{MEs} and \r{ME-IsCh}. We started with the result \r{mat-el} obtained
in our previous paper \cite{gipst1} using the Sklyanin method of Separation of Variables by which we obtained explicit
wave functions in terms of the solutions of Baxter equations. The result \r{mat-el} was obtained for the
general $N=2$ BBS-model which is related to a generalized free-fermion Ising-type model \r{FFIsing}. For this
general model we were able to get the much simpler formula \r{mat-res}
which, however still involves a multiple sum over intermediate spins.
Since for the general model we cannot
perform the summation, for the further calculation we restricted ourselves to the Ising model
with parameters \r{Isi}. Performing a number of technical manipulations, we succeed in calculating the multiple
spin sums explicitly. Although the intermediate formulas are quite involved, a number of surprising cancellations take
place which lead to the rather simple formula \r{ME} for the spin matrix element square. This comes still normalized
to the auxiliary states involved in the method of Separation of Variables, but it is not difficult to convert this
result into the properly normalized matrix elements for the model with periodic boundary condition.
The final formula becomes more lengthy
due to normalization factors. We show by which transformations we get the recently conjectured formula of
A.~Bugrij and O.~Lisovyy. Our derivation provides a first proof of these formulas.
Another application of the formulas obtained in this article is the result \r{ME-IsCh} for
the spin matrix elements for the finite quantum Ising chain in a transverse field.

The presence of degenerations in the spectrum for
the special Ising parameter values forced us in this case to normalize the Baxter equation solutions
differently for different
excited states. The complexity of the formulas gives little hope that for more general parameter values
the multiple spin summations can be done in the near future, even if then the degeneration problems may be avoided.

\ack
The authors thank A.~Bugrij and O.~Lisovyy for helpful discussions.
This work has been supported by the Heisenberg--Landau exchange program HLP-2007.
SP was supported by the RFBR grant 06-02-17383 and the grant for support of scientific schools
NSh-8065.2006.2. NI and VS  were supported by the INTAS grant 05-1000008-7865,
by the Program of Fundamental Research of the Physics and
Astronomy Division of the NAS of Ukraine and the Ukrainian DFFD grant
$\Phi 16$-457-2007.
The research of YT was supported in part by the Eiffel Fellowship of the French Government.

\setcounter{section}{0}
\setcounter{subsection}{0}

\renewcommand{\thesection}{\Alph{section}}

\section{Proof of the summation formula}

The aim of this Appendix is to find a factorized expression for the sum
\be\label{YD}Y_{\CD}\,
 \;=\;\sum_{\rho_{l},\,\,l\in\CD}\; \frac{\prod_{l\in\CD}\; (-1)^{\rho_{l}}\;
     ((-1)^{\rho_{l}} r_{l}\,+\, \xi_l)
\;\:((-1)^{\rho_{l}} r_{l}\, +\fba)} {\prod_{l<m,\: l,m\in\CD}\;
     ((-1)^{\rho_{l}} r_{l}\:+\, (-1)^{\rho_{m}}r_{m})}
\ee
where $\xi_l$ is given by \r{xi} with \r{SSF}.
The result for $\xi_l=\pm\alpha_{\tilde q_l}$ is
\bea \fl Y_{\CD}
 &=&\sum_{\rho_{l},\,\,l\in\CD}\; \frac{\prod_{l\in\CD}\; (-1)^{\rho_{l}}\;
     ((-1)^{\rho_{l}} r_{l}\,\pm\, \alpha_{\tilde q_l})
\;\:((-1)^{\rho_{l}} r_{l}\, +\,\fba)} {\prod_{l<m,\: l,m\in\CD}\;
     ((-1)^{\rho_{l}} r_{l}\:+\, (-1)^{\rho_{m}}r_{m})}\ny\\
\fl &=& \lb
 c_D^{\rm odd}\,(b\,\pm\,a)\lk \prod_{j\in\CD}\qq{j}\,\mp\,a\,b\rk\:
    {\displaystyle \frac{\prod_{l\in\CD}\: (2\,r_l/a)
  \:f_l^{(D-1)/2}\;g_l^{(D-3)/2}}{\prod_{l,m\in\CD,\,l<m}\;(\pm h_{l,m})}}\quad D\;{\rm odd}
\atop
     {c_D^{\rm even}\,(a \pm b)\lk a \prod_{j\in\CD}\qq{j}\:\mp b\rk\;
   {\displaystyle \frac{\prod_{l\in\CD}\:(2\,r_l/a)\,(f_l\,g_l)^{D/2-1}}
    {\prod_{l,m\in\CD,\,l<m}\:(\pm h_{l,m})}}\ \ \qquad\;\: D\;{\rm even}}
  \,\right.
\label{SUM}\eea
with $D=|{\cal D}|$,
\be\label{cab}\fl\qquad c_D^{\rm odd}\:=\:\aab^{-(D-1)(D-3)/4}\:(-\abb)^{-(D-1)^2/4}\,,\hx
     c_D^{\rm even}\:=\: \aab^{-(D-2)^2/4}\:(-\abb)^{-(D-2)D/4}\,,\ee
and we use the abbreviations \r{fgh}.
Replacing $\pm\alpha_{\tilde q_l}\rightarrow\pm\beta_{\tilde q_l}\;$ in \r{SUM}, see \r{SSF}, \r{xi}, amounts to
$\;\aab\leftrightarrow\abb\;$ in $\;c_D^{\rm odd}\;$ and $c_D^{\rm even}$, and
\bea
   (b\pm a)\,(\,\pj\mp a\,b)\;&\rightarrow&\; (1\mp a\,b)\,(\:b\:\pj\,\pm\, a)\qquad
  \quad D\;{\rm odd},
  \ny\\[2mm]
(a\pm b)\,(a\,\pj\,\mp b)\;&\rightarrow&\; (1\mp a\,b)\,(\pm a\,b\,\pj+1)\qquad
D\;{\rm even} .\ny
\eea

In Section \r{summ} the last bracket in the numerator of \r{YD}
is also needed with $+\fba=b/a$ replaced by $-\fba$. However, in order not to complicate the formulas, we
shall always use \r{YD} as it is written here. Since the only other $b$-dependence, which is in $\xi_j$ and $r_j$,
is quadratic,
this can be adjusted at the end of the calculation just by changing the sign of $b$.
Also, it is sufficient to prove formula \r{SUM} for the case $\xi_l=\alpha_{\tilde q_l}$.
Then the results for the cases
$\xi_l=-\alpha_{\tilde q_l}$ and $\xi_l=\pm \beta_{\tilde q_l}$ can be obtained by simple transformations of
the variables $r_l$ and $b$.

First, we find the recurrence relation for $Y_{\mathcal{D}}$ with respect to
$D=|\mathcal{D}|$.
It relates this quantity for $D$ and $D-2$.
Then we verify \r{SUM} for small $D$.
Finally,
by explicitly inserting our conjectured solutions \r{SUM} into the recursion relation we
prove the expression. The cases of $D$ odd and even have
to be treated separately, since we are dealing with a two-step relation.

\subsection{Derivation of the recursion relation}
This recurrence relation for $Y_{\cal D}$
is obtained by using the identity, compare \r{sumru},
\be\label{inter}
\sum_i \frac{\prod_k(x_i-y_k)}{\prod_{j\ne i}(x_i-x_j)} = 0\ ,
\ee
when the number  of $x_i$ exceeds the number of $y_k$ at least  by two.
Fix any index $s\in\mathcal{D}$
 and consider $D+1$ values of $x_i$ and two values of $y_i$:
 $\{y_1, y_2\}=\{-\xi_s, -\fba\}$,
\[\fl  \{x_0,x_1,\ldots,x_{D}\}=\{r_s,-r_s,-(-1)^{\rho_1}r_1,-(-1)^{\rho_2}r_2,\ldots,
         \underbrace{-(-1)^{\rho_s}r_s}\limits_{\rm omitted},\ldots,-(-1)^{\rho_{D}}r_{D}\}.\]
Since two parameters $y_k$ are chosen, we must have $D\geq 3$. Now we separate the two terms in \r{inter}
which correspond to $i=0,1$ and present them as a summation over $\rho_s\in\{0,1\}$ for
$\{x_0,x_1\}=r_s(-1)^{\rho_s}$.
Then \r{inter} becomes
\bea
\lefteqn{\sum_{\rho_s} \frac{(-1)^{\rho_s}(r_s(-1)^{\rho_s}+\xi_s)(r_s
(-1)^{\rho_s}\,+\,\fba)}{\prod_{k\ne
s}(r_s(-1)^{\rho_s}+r_k(-1)^{\rho_k})} }\ny\\
&&\hs\hs=\; -\sum_{k\ne s}\frac{2\,r_s
(-r_k(-1)^{\rho_k}+\xi_s)(-r_k(-1)^{\rho_k}\,+\,\fba)}{(r_k^2-r_s^2)\prod_{l\ne
k,s}(-r_k(-1)^{\rho_k}+r_l(-1)^{\rho_l})}\ .
\label{identi}\eea
Now in \r{YD} we separate the summation over a certain fixed discrete variable $\rho_s$ and use the identity \r{identi}
to replace the summation over $\rho_s\in\{0,1\}$ by a summation over $k$.
After this, we move the summation over $k$ to the front of formula \r{YD}
and collect the factors depending on $\rho_k$:
\bea \fl \lefteqn{Y_{\mathcal{D}} = -\sum_{k\neq s}\;\sum_{\rho_{l\in \mathcal{D}/{s,k}}}
\frac{\prod_{l\neq s,k}(-1)^{\rho_l}(r_l(-1)^{\rho_l}+\xi_l)(r_l(-1)^{\rho_l}+\,\fba)}
{\prod_{l<m,l,m\neq s,k}(r_l(-1)^{\rho_l}+r_m(-1)^{\rho_m})}\;\times}
\label{YDb}\\ \fl &\:\times&\sum_{\rho_k}\,\frac{(-1)^{\rho_k}(r_k(-1)^{\rho_k}+\xi_k)(r_k(-1)^{\rho_k}+\,\fba)}
{\prod_{l\neq k,s}(r_k(-1)^{\rho_k}+r_l(-1)^{\rho_l})}\;\frac{2r_s(-r_k(-1)^{\rho_k}+\xi_s)(-r_k(-1)^{\rho_k}+\,\fba)}
{(r_k^2-r_s^2)\:\prod_{l\neq k,s}(-r_k(-1)^{\rho_k}+r_l(-1)^{\rho_l})}\,.
\ny\eea
Now after multiplication in the second line of \r{YDb},
we perform the summation over $\rho_k\in\{0,1\}$ by means of the relation
\[\textstyle
\sum_{\rho_k}\,(-1)^{\rho_k}(r_k(-1)^{\rho_k}+\xi_k)(-r_k(-1)^{\rho_k}+\xi_s)=-2r_k(\xi_k-\xi_s)\,.
\]
Finally, we get the desired recursion relation, valid for any fixed index $s\in\CD$:
\be\label{rec1}
Y_{\mathcal{D}}\:=\: \sum_{k\ne s}\;Y_{\mathcal{D}/k,s}\;\frac{ 4\,r_k\,r_s}{(r_k^2\,-\,r_s^2)}
  \,\frac{(\fba^2\,-\,r_k^2)\;(\xi_k\,-\,\xi_s)}
        {\prod_{l\ne k,s}(r_l^2\,-\,r_k^2)}\,,\ee
where $Y_{\mathcal{D}/k,s}$ is just the original $Y_{\mathcal{D}}$
with the indices $k$ and $s$ removed from $\CD$.

Since \r{rec1} is a two step difference equation, in general it has two independent solutions
 $Y^{(1)}_\CD\:$ and $Y^{(2)}_\CD\,$.
The quantity \r{YD} is equal to a linear combination
of these solutions, $Y_\CD\,=\,c_1\,Y^{(1)}_\CD\,+\,c_2\,Y^{(2)}_\CD$
and the constants $c_1$, $c_2$ are fixed by calculating $Y_\CD$ explicitly for small $D$.

In the following, using \r{rec1}, we give the details of the proof of \r{SUM} for the case $D$ odd.
The derivation for $D$ even proceeds analogously.

\subsection{Proving the summation formula for $D$ odd.}

For $D=1$ we start calculating directly \r{YD}, $\alpha_s=\alpha_{\tilde q_s}$:
\[\fl
Y_{\{s\}}=\sum_{\rho_s}(-1)^{\rho_s}(r_s(-1)^{\rho_s}+\alpha_s)(r_s(-1)^{\rho_s}+b/a)
=2 r_s\frac{(b+a) (e^{{\rm i}\tilde q_s}-a b)}{a(e^{{\rm i}\tilde q_s}-a^2)}\, ,
\]
which proves \r{SUM} for this particular case.
This result, together with the recursion relation \r{rec1}, defines
$Y_{\CD}$ for odd $D$ uniquely.
For example, for the set $\CD=\{s_1,s_2,s_3\}$, the recursion relation
\r{rec1} with selected $s=s_3$ gives
\bea\fl  Y_{\lb s_1,s_2,s_3\rb}\;=\;Y_{\lb s_2\rb}\frac{4r_1r_3}{(r_1^2-r_3^2)}
   \frac{(\fba^2-r_1^2)(\alpha_1-\alpha_3)}{(r_2^2-r_1^2)}
   +Y_{\lb s_1\rb}\frac{4r_2r_3}{(r_2^2-r_3^2)}
   \frac{(\fba^2-r_2^2)(\alpha_2-\alpha_3)}{(r_1^2-r_2^2)}
  \ny\\
   \fl =\;
  \frac{(b+a)\prod_{l=1}^3\,(2r_lf_l)}{a^3\,(-\abb)\,
  \prod_{1\le l<m\le 3} h_{l,m}}\lk \frac{\qq{1}(\qq{2}-ab)h_{2,3}}{(\qq{2}-\qq{1})}
   +\frac{\qq{2}(\qq{1}-ab)h_{1,3}}{(\qq{1}-\qq{2})}\right)\ny
   \\
   =\;
\frac{(b+a)\prod_{l=1}^3\,(2r_lf_l)}{a^3\,(-\abb) \prod_{l<m}h_{l,m}}\;
\left({\textstyle \prod_{l=1}^3}\,\qq{l}-ab\right)\,, \label{reev}
\eea
where we used
\be  \alpha_k-\alpha_s=-\aab\,\rho_{k,s},\hx\;
  \fba^2\!-\:r_k^2=\frac{\aab\,\abb\,\qq{k}}{a^2\,f_k\,g_k},\hx\;
  r_l^2-r_k^2=\frac{\aab\,\abb\,h_{l,k}}{f_l\,f_k}\;\rho_{l,k} \label{aer}
  \ee
with $\rho_{k,s}\,=\,(e^{{\rm i}\tilde q_k}-e^{{\rm i}\tilde q_s})/(g_k\,g_s)$.
Observe that the big bracket in the second line of \r{reev} factorizes and leads to a result
symmetrical in the three indices. The result obtained in \r{reev} proves
\r{SUM} for $D=3$ and $\xi_l=\alpha_{\tilde q_l}$.
We can easily continue this recursive procedure to conjecture
the formula \r{SUM} for odd $D$. To prove it, it is enough to show that
the right-hand side of \r{SUM} satisfies the recursion relation \r{rec1}.
The right-hand side of \r{SUM} can be presented as
$Y_{\mathcal{D}}=c_1Y^{(1)}_{\mathcal{D}}+c_2Y^{(2)}_{\mathcal{D}}$, where
$c_1=-(a+b)ab$, $c_2=a+b$ and
\begin{equation}
 Y^{(1)}_{\mathcal{D}} = c_D^{\rm odd}\, \frac{\prod_l (2 r_l/a) f_l^{(D-1)/2} g_l^{(D-3)/2}}
{\prod_{l<m}h_{l,m}}\ ,\quad Y^{(2)}_{\mathcal{D}} =Y^{(1)}_{\mathcal{D}}
\prod_l e^{{\rm i}\tilde q_l}\,.
\label{YD1}
\end{equation}
Thus to prove \r{SUM}, it suffices to prove that $Y^{(1)}_{\mathcal{D}}$ and $Y^{(2)}_{\mathcal{D}}$
satisfy \r{rec1}.

Let us prove that $Y^{(1)}_{\mathcal{D}}$
satisfies the recurrence relation \r{rec1} with $\xi_l=\alpha_{\tilde q_l}$.
Using the elementary relation $c_{D-2}^{\rm odd}/c_D^{\rm odd}=-\aab^{D-3}\abb^{D-2}$ and
\r{aer},
we reduce the problem to the proof of the identity
\bea \fl \lefteqn{\frac{\prod_l (2r_l/a)\:f_l^{(D-1)/2}\:g_l^{(D-3)/2}}
{\prod_{l<m}h_{l,m}}\;=\;
\sum_{k\neq s}\;\frac{\prod_{l\neq k,s} (2r_l/a)\:f_l^{(D-3)/2}g_l^{(D-5)/2}}
{\prod_{l<m;\,l,m\neq k,s}\:h_{l,m}}\:\:\times}\ny\\
         &&\hs\times \:\frac{4\;r_k\,r_s}{a^2}\;\:
         \frac{f_s\,f_k^{D-2}\,g_k^{D-3}
        \qq{k}}{h_{s,k}}
        \prod_{l\neq k,s}\frac{f_l\,g_l}{h_{l,k}\:(\qq{l}-\qq{k})}\,.
  \label{matc} \eea
We see that all $r_l$ and the $f_l,g_l,h_{l,m}$ with
     $l,m\neq k,s$ match between both sides of \r{matc}, leaving us with
\be \frac{(f_s\,g_s)^{(D-3)/2}}{\prod_{l\neq s}\:h_{l,s}}\;=\;
  \sum_{k\neq s}\;\frac{(f_k\,g_k)^{(D-3)/2}\;e^{{\rm i}\tilde q_k}}{h_{k,s}\;
  \prod_{l\neq s,k}\:(e^{{\rm i}\tilde q_l}-e^{{\rm i}\tilde q_k})}\,,\label{recc}\ee
which is equivalent to the interpolation identity \r{inter} used here
for the following choice of the parameters:
$x_k= \qq{k}$ for $k\in\mathcal{D}/s$, $x_s = e^{-{\rm i}\tilde q_s}$,
\[\{y_0,y_1,\ldots,y_{D-3}\}=
\{0,\underbrace{a^2,\ldots,a^2}\limits_{(D-3)/2\, \mbox{\footnotesize{times}}},
\underbrace{a^{-2},\ldots,a^{-2}}\limits_{(D-3)/2\, \mbox{\footnotesize{times}}}\}\,.
\]
The proof that $Y^{(2)}_{\mathcal{D}}$
satisfies the recurrence relation \r{rec1} reduces to the same interpolation identity,
but with $y_0$ omitted.
This proves \r{SUM} for $D$ odd and $\xi_l=\alpha_{\tilde q_l}$.

As was explained before, the change of the sign at $\zeta$ in \r{YD} can be adjusted by changing the sign of $b$
in the final formula. Similarly, the change of the sign at $\xi_l$ can be adjusted by the simultaneous change of the
the signs of $r_l$ and $\zeta$. Finally, the transformation $b\to 1/b$ leads to $r_l\to r_l/b^2$,
$-\alpha_{\tilde q_l}\to \beta_{\tilde q_l}/b^2$. This allows to find \r{YD} at $\xi_l=\beta_{\tilde q_l}$
if we know \r{YD} at $\xi_l=-\alpha_{\tilde q_l}$.
The mentioned transformations cover all the cases of \r{YD} for $D$ odd.

\subsection*{References}
\bibliographystyle{amsplain}

\end{document}